\documentclass[twocolumn,times,tighten]{aastex63}
\usepackage{natbib}
\accepted{by ApJ, 2020 Aug 21}
\shorttitle{The Central 300\,pc of the Galaxy probed by H$_3^+$
  and CO} \shortauthors{Oka et al.}
\begin{document}

\title{ The Central 300\,pc of the Galaxy probed by infrared spectra
  of H$_3^+$ and CO: II. \\ Expansion and morphology of the warm
  diffuse gas}

%------------------------------------------------------------
% \correspondingauthor{Takeshi Oka}
\author{Takeshi Oka}
%\author[0000-0000-0000-0000]{Takeshi Oka}
\affiliation{Department of Astronomy and Astrophysics and
  Department of Chemistry, the Enrico Fermi Institute,
  University of Chicago, Chicago, IL 60637 USA;
  t-oka@uchicago.edu}
% \email{t-oka@uchicago.edu}
\author{T. R. Geballe} \affiliation{Gemini Observatory, Hilo, HI 96720 USA}
\%================================================================
\vspace{4mm}
\begin{abstract}
  
Velocity profiles of a line of H$_3^+$ at 3.7\,$\mu$m produced in warm
diffuse gas have been observed toward 18 stars in the Central
Molecular Zone (CMZ) of the Galaxy. Their longitude-velocity diagram
indicates that the gas is radially expanding within the CMZ at speeds
up to a maximum of $\sim$150\,km\,s$^{-1}$. The current momentum and energy in the 
gas are $\sim 5 \times 10^8~M_\odot$\,km\,s$^{-1}$ and $\sim 5\times 10^{53}$\,erg. 
The motion is similar to that of the Expanding Molecular Ring (EMR) discovered in 1972 by Kaifu 
et al. and by Scoville. We propose that the expanding gas seen in H$_3^+$ is part 
of the same phenomenon, in spite of differences in estimates of density, morphology, and 
degree of rotation. The outward motion suggests that one or more ejection events occurred 
near the center of the CMZ (0.5--1) $\times$ 10$^6$\,years ago, which may be related to 
creation of the recently observed microwave bubble. These observations revive the circular 
face-on view of the CMZ proposed in 1972, which fell out of favor after 1991 when Binney et
al. proposed that a face-on view of the CMZ would show it to
have an elliptical shape, with high eccentricity. While that model may
apply on kiloparsec scales, we argue that it is incorrect to apply it
to the much smaller CMZ. We discuss the fate of the expanding gas,
which appears to be eventual infall into the center, leading to
episodes of star formation and violent events associated with
accretion onto Sgr~A$^\ast$.

\end{abstract}
%================================================================
\section{INTRODUCTION}

On the night of July 10-11, 1997 at the United Kingdom Infrared
Telescope (UKIRT) spectra toward two bright stars in the Galaxy's 
Central Molecular Zone (CMZ), the central 300\,pc of the Galaxy,
revealed unexpectedly high column densities of H$_3^+$ \citep{geb99}. 
One of the stars, GCIRS~3, is in the Central Cluster of
massive luminous stars centered on the supermassive black hole,
Sgr~A$^\ast$. The other, known as GCS~3-2, is located in the Quintuplet 
Cluster, 30\,pc to the east of Sgr~A$^\ast$, and is also deeply embedded 
in the CMZ. The column density of H$_3^+$ toward each of these stars is 
an order of magnitude higher than the column densities of H$_3^+$ that 
had been measured toward young stellar objects in dense clouds and toward 
a star in a diffuse cloud, all of which are located in the Galactic disk 
\citep{geb96, mcc98, mcc99}. This striking difference exists despite the extinctions 
to the two stars in the GC stars being comparable to and in some 
cases lower than the extinctions to the young stellar objects, and only three times 
higher than the extinction to the diffuse cloud source. Consequently the
3.5--4.0\,$\mu$m infrared spectrum of H$_3^+$ \citep{oka80}, or
trihydrium, a molecular ion having several unique and valuable
properties for studies of the interstellar medium, has emerged as a
novel probe to investigate the gaseous environment in the CMZ.

Further observations and analysis in the first decade of this century
established that the CMZ in the region from Sgr~A$^\ast$ to 30\,pc east
contains long columns of warm ($\sim$200\,K) low density
($\lesssim$100~cm$^{-3}$), and mostly blueshifted gas \citep{got02, oka05, got08}. 
In 2008 a program was launched to
search for bright stars throughout the CMZ with smooth continua in
order to extend studies of H$_3^+$ from the central 30\,pc to
cover the full 300\,pc extent of the CMZ \citep{geb10}. The
project has been completed \citep{geb19a} and has provided
$\sim$30 suitable stars extending from 140\,pc west of Sgr~A$^\ast$ to
116\,pc east (herein we use a GC distance of 8\,kpc). A
comprehensive analysis of spectra of H$_3^+$ and CO observed during
the last 20 years, including many of these newly found objects, has
recently been published \citep[][hereafter Paper~I]{oka19}. 
This has led to two conclusions: (1) the predominance in terms of volume of 
warm and diffuse gas in the CMZ; and (2) the high cosmic ray ionization rate
($\zeta \sim 10^{-14}$\,s$^{-1}$) in the CMZ, 1000 times higher in
dense clouds \citep{geb96, mcc99} and 100 times
higher than in diffuse clouds in the Galactic disk \citep{mcc03, ind07, ind12}.
As corollaries, Paper~I showed that the CMZ is not as opaque as reported in some papers
\citep[e.g.][]{mor96} and concluded that ultra-hot
X-ray-emitting gas does not exist continuously and extensively in the
CMZ as reported in some papers \citep[][see their Fig.~9]{koy89,laz98} .
   
In this paper, Paper~II, we report the results of our investigations
of the dynamics and morphology of this warm and low density gaseous
environment, obtained from the velocity profiles of H$_3^+$ lines. We
take advantage of properties of the absorption profiles produced in
dense clouds within the CMZ and the foreground spiral arms: that they
tend to be sharp and occur at radial velocities known from previous
radio and millimeter wave spectroscopy. In contrast, the profiles of
H$_3^+$ lines arising within the CMZ, in addition to frequently being
broad, occur predominantly in warm diffuse gas, an environment that is
usually straightforward to differentiate from gas in dense clouds via
spectroscopy of several H$_3^+$ lines and lines of the first overtone
band of CO. The velocity dispersions of the H$_3^+$ lines arising in
the warm diffuse gas along the line of sight are as high as
150\,km\,s$^{-1}$, a unique feature of the CMZ. We make use of the
longitude-velocity ($l$, $v$) diagram \citep[e.g.][Section 9.1]{bin98}
of this gas to draw simple and straightforward
conclusions regarding its dynamics and morphology.

%============================================================
\section{PREVIOUS OBSERVATIONS AND INTERPRETATIONS} % section 2
%------------------------------------------------------------
\subsection{Early Observations} % section 2.1

The highly blueshifted line profiles that play a central role in this
paper were first observed in the 21\,cm H{\sc i} emission spectrum by
\cite{rou60}. Due to the very low spatial resolution, the
observed maximum positive and negative velocities of 135\,km\,s$^{-1}$
were interpreted by them as the maximum tangential velocity of a
circular motion (see Fig.~4 of \citet{rou60}, Fig.~1 of \citet{rou64}, and Fig.~9 of 
\citet{oor77}). However, based on a separate survey of high velocity H{\sc i} 
emission \citet{kru70} had proposed an expulsion of gas from 
the Galactic nucleus.

When molecular radio astronomy was initiated with observation of the
18\,cm $\Lambda$-doublet OH absorptions \citep{wei63}, five
papers were published in 1964 on OH absorption in the GC using Sgr~A
as the background radio continuum source \citep{bol64a,die64,rob64,gol64,bol64b}
primarily because it is bright. Among them Fig.~2 of
\citet{rob64} most clearly demonstrated the existence of the
$-$135\,km\,s$^{-1}$ absorption toward Sgr~A. Readers are referred to
\citet{whi94} for more details of the spectra and interesting
anecdotes of the Australian OH observations. Unlike the H{\sc i}
emission, which does not allow one to determine if the high velocity
gas is situated in front of or behind Sgr~A, the intense OH absorption
demonstrated that a large amount of gas is moving outward from the GC
at high velocity. Later \citet{rob70} and \citet{mcg70} 
reported their extensive survey of one of the $\Lambda$-doublet lines at
1667\,MHz from Galactic longitude 357\degr30\arcmin~to
3\degr20\arcmin, measuring characteristic radial velocities, velocity
dispersions, sizes, and maximum values of apparent opacities for 63
clouds.

%------------------------------------------------------------
\subsection{The Expanding Molecular Ring (EMR)}
% 8\,kpc to GC is mentioned for the second time after INTRODUCTIon

Based on the ($l$, $v$) diagram of the radio OH absorption in Fig.~4
of \citet{mcg70} obtained using the 64\,m Parkes Radio Telescope
(angular resolution 12\farcm2), \citet{kai72} proposed the
existence of an Expanding Molecular Ring (EMR), a circular ring of
radius 220\,pc (adjusted to the GC distance of 8\,kpc used herein)
with expansion and tangential velocities of 130\,km\,s$^{-1}$ and
50\,km\,s$^{-1}$, respectively; observations of the rear part of the ring 
were supplemented by measurements of NH$_3$ emission at two positions. 
Kaifu et al. also pointed out the existence of the ring in the ($l$, $v$) 
diagram of radio H$_2$CO absorption in Fig.~3a of \citet{sco72a}
albeit with less angular coverage.

In a paper received 16 days after \citet{kai72}, \citet{sco72b},
using the 43\,m Green Bank Telescope, also proposed the existence
of the EMR, based on the ($l$, $v$) diagram of an H$_2$CO absorption
line, but with a somewhat smaller radius 170\,pc (adjusted) and 
expanding and tangential velocities of 145\,km\,s$^{-1}$ and 50\,km\,s$^{-1}$,
respectively. These values are taken from Model~I of Scoville's
Table~I; we ignore Model~II (Scoville also mentions that ``contraction
cannot be ruled out''). According to both papers the EMR is situated in the
Galactic plane. In estimating the total mass of the EMR both papers assumed
gas densities on the order of 10$^3$--10$^4$\,cm$^{-3}$;
it is now thought that the observed gas is more likely to be
mostly lower density (see Section 5.3.1).

Two years later \citet{kai74} reported ``an almost complete
ring'' from observations of H{\sc i}\,21\,cm absorption in front of
Sgr~A and emission behind it. We show the inferred face-on views
of the CMZ of \citet{kai72} and \citet{sco72b} in
Figure~1, because the circular shape of the CMZ and expansion at
$\sim$140\,km\,s$^{-1}$ are similar to what we conclude in this 
paper on the morphology and motion of the warm and diffuse gas within 
the CMZ. Note that the geometry and expansion of the EMR were derived from 
absorption spectra. Unlike the motions of dense gas in the GC, which have been
explained as responses to the gravitational potential (see the next section), 
there is no way to explain the EMR in that way. (See \citet{lis80}, 
however, for an explanation of a larger scale expanding arm as the result
of the gravitational potential.) The motion of the EMR must be due to a massive 
expulsion of gas in the relatively recent past (see Section~6.2).

For a recent 3-dimensional analysis of the EMR as an expanding molecular 
shell \citep{sof95b} and an expanding molecular cylinder \citep{sof17} based 
on radio CO emission, see the next subsection.

%============================================================
% Figure 1
%============================================================
\setcounter{figure}{0}
\begin{figure*}
\includegraphics[angle=-0,width=0.98\textwidth]{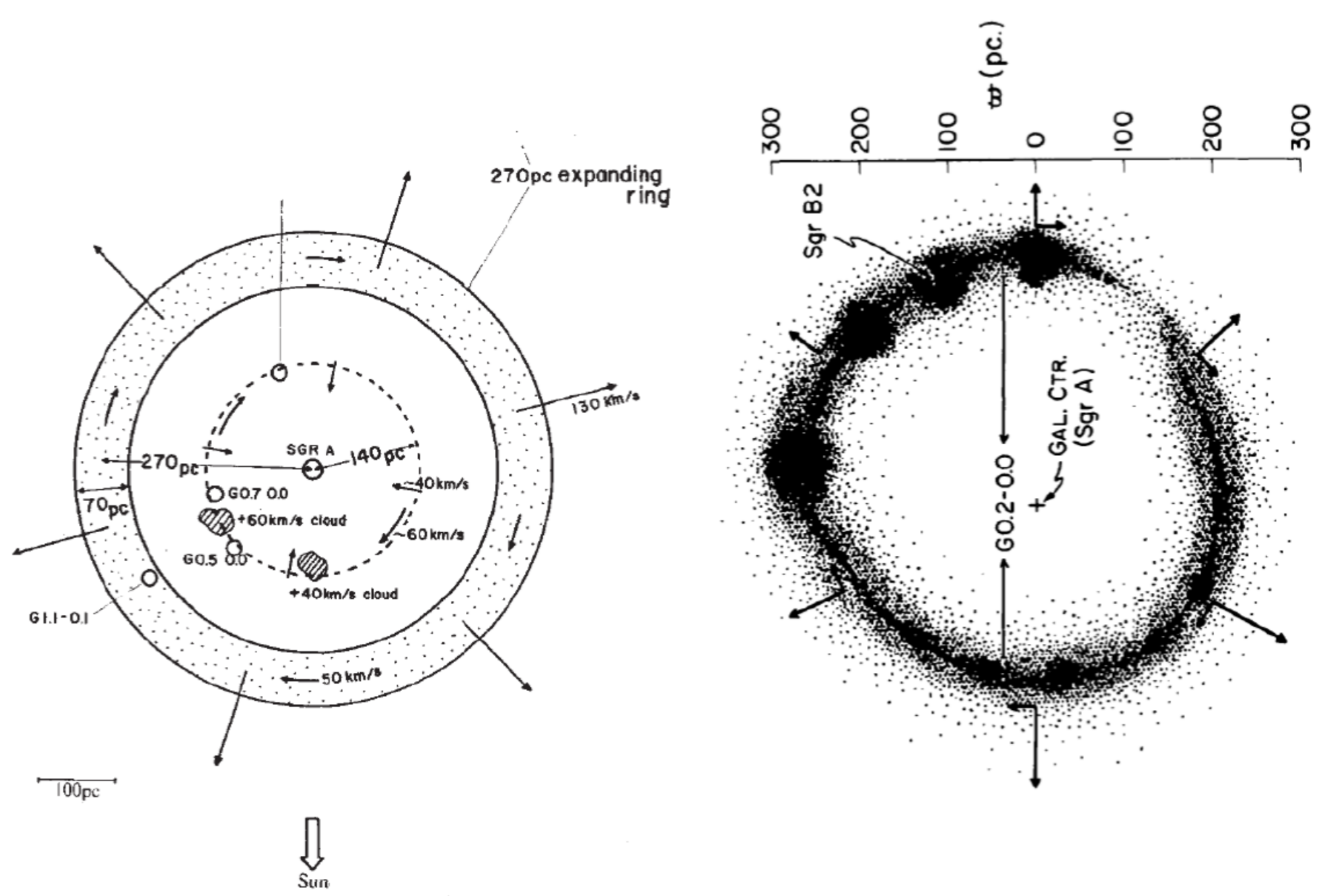}
\caption{Face-on views of the Expanding Molecular Ring from 
\citet{kai72} (left) and \citet{sco72b} (right).
Note that the scales of the rings in pc in these figures are based 
on a GC distance of 10~kpc.}
\end{figure*}
%============================================================

%------------------------------------------------------------
\subsection{The Galactic Center Ring (GCR)} % subsection 2.3

Many molecules have been used to produce ($l$, $v$) diagrams
subsequent to the proposed existence of the EMR. By far the one most
extensively used is the CO molecule, whose radio emission arises virtually entirely in dense gas 
\citep{ban77,lis78,hei87,bal87,bal88,sof95a,sof95b,dah97,oka98a,oka98b,saw01,mar04,oka12}.
The slanted nearly linear section of the ($l$, $v$) diagram visible in Fig.~1e of 
\citet{hei87}, Fig.~4 (especially at $b=0.0$) of \citet{bal87}, and Fig.~3 
of \citet{lis92} indicate the presence of a rotating ring of dense gas. This was isolated in a
detailed analysis of the data in \citet{bal87} by \citet{sof95a}
and was called ``the 120-pc molecular ring'' (110\,pc for the GC
distance used in this paper). This ring of purely rotating dense
clouds tilted by 5\degr~and slightly bent has been widely accepted and
named the Galactic center ring (GCR) by \citet{rod06}. The 
``100\,pc ring'' observed in far-infrared emission by
dust \citep{mol11} may be the same structure, although there are
significant morphological differences. The active star forming regions
Sgr~B2 and Sgr~C are often placed on the GCR.
 
Therefore, there appeared to be two rings, the outer EMR with a radius
of $\sim$200\,pc which suggested some explosive event near the Galactic
nucleus within a million years (see Section~6), and the inner
rotating circular GCR of dense gas with a radius of $\sim$100\,pc
which is the response to the central gravitational potential. Subsequently 
the GCR has been re-interpreted as composed of two spiral arms, 
Arm I and Arm II (see Figs. 2 and 3 of \citet{sof95a} and Fig. 3 of \citet{sof95b}, 
\citet{saw04}, \citet{hen16}, \citet{rid17}, and others).

Sofue's three-dimensional analysis of the CO emission observed by \citet{bal87, bal88} also significantly 
altered the concept of the EMR \citep{sof95b}. Instead of a planar structure he found a structure with 
vertical extent of $\pm$50 pc and called it the Expanding Molecular Shell (EMS). More recently, from 
a three-dimensional analysis of more extensive observations by \citet{oka98b}, \citet{sof17}
showed that the EMR is a bi-polar vertical cylinder with the total length as great as 170 pc 
and called it the Expanding Molecular Cylinder (EMC). He also identified the GCR with the
CMZ and estimated that the mass of the CMZ is higher than that of EMC by a factor of $\sim$8. 
His estimate of the mass of the EMR/EMC of 0.8 $\times$ 10$^7$ M$_\odot$ is in good agreement
with the mass of 6 $\times$ 10$^6 M_\odot$ (Table 7 of Paper I). See also \citet{hen16}.

%------------------------------------------------------------
\subsection{Barred potential and elliptical orbit with high eccentricity}
% subsection 2.4

The interpretation of the gas in the CMZ as having a largely circular distribution, 
as suggested by the EMR and GCR, was drastically modified by \citet{bin91}, 
who developed a theory of gas motion in the Galactic barred potential 
\citep[see also][] {ana92a, ana92b}. They demonstrated that for longitudes
within 10\degr~($\sim$1,500\,pc) of the center the motions of dense
clouds observed in mm wavelength lines of CO and CS can be accounted
for by two closed orbits, $x_1$ and $x_2$, so named by \citet{con77} 
in their non-linear theory of inner Lindblad
resonances in galaxies (\citet{lin27}; see also \citet{con75}). 
The $x_2$ orbit is circular and hence its ($l$, $v$) diagram is
a straight line while the $x_1$ orbit is an ellipse of high
eccentricity, with sharp cusps at the two ends and its ($l$, $v$)
diagram resembles a parallelogram \citep[see Figs.~1 and 3 of][]{bin91}. 
The theoretical calculation appears to reproduce the large
scale ($l$, $v$) diagram of the 21~cm\,H{\sc i} emission observed by
\citet{bur78} in which the longitudinal coverage is
$\sim$3\,kpc. There exist some similarities between the calculated
$x_1$ orbit and observed face-on views of external galaxies
\citep[e.g.,][]{reg99}. The ellipse is seen clearly in the observed
face-on views of external galaxies with bars within bars, e.g., 
Fig.~1a-d of \citet{fri93} and Fig.~14 of \citet{kor04}. 
The $x_1$ orbit applied to gas on the scales of
kiloparsecs seems to be well supported by observations. Many additional
studies of the Lindblad resonance applied to large scale GC gas distribution 
and kinematics have been published  \citep[e.g.][]{sor15a,sor15b,sor15c,suz15}.

However, application of the theory in Sections 3, 4, and 5 of \citet{bin91} 
to the more than 10 times smaller volume of the CMZ is
questionable. Unlike on kiloparsec scales there is no observational
evidence in the CMZ for the existence of a barred potential  \citep{kru15}. 
Therefore there is no reason for the existence of an
ellipse with high eccentricity which surrounds a bar. In the nuclear
spirals model of \citet{rid17}, based on the two spiral arms
reported by \citet{sof95a} and \citet{saw04}, the $x_1$ orbit is
introduced {\it a priori} as connecting to Arm I of \citet{sof95a},
but the source of the barred potential causing the $x_1$ orbit is not
identified. We conclude that the resemblance of the ($l$, $v$) diagram
of CO emission in the CMZ to an off-center parallelogram in Fig.~2
of \citet{bin91} (from \citet{bal88}; see also \citet{bli93})
must be incidental and that the face-on view of the CMZ as
an ellipse with extremely high eccentricity, which appears for example
in Fig. 2 of \citet{rod06}, Fig.~21 of \citet{bal10}, and Fig.~4 of \citet{tsu18},
is a chimera. Nevertheless the elliptical $x_1$ orbit has become popular
and the notions of an EMR and the face-on view of the CMZ being
circular seem to have taken a back seat. Few papers have been
published on the EMR after \citet{sof95b}. The ``EMR'' in Fig.~12 of
\citet{saw04} is a misnomer since the gas motion is tangential
to the ring. Our analysis of the observations presented in this paper 
revives the circular face-on view of the CMZ shown in Fig.~1.

%------------------------------------------------------------
\subsection{H$_3^+$ as a probe for dynamical and morphological studies of the CMZ} % subsection 2.5

The most extensive spectroscopic survey of gas in the CMZ has been
that of the CO\,$J=1\rightarrow 0$ emission line by \citet{oka98b}
(Tomoharu Oka of Keio University, not to be confused with the first
author of the present paper). Emission from this line nearly
completely covers the CMZ. In comparison, the high-resolution H$_3^+$
absorption spectroscopy of CMZ gas reported here has the serious
drawback that observations are limited to sightlines to bright
stars with smooth infrared continua \citep{geb19a}. As
mentioned in the Introduction, such stars are rare. Moreover, while
the 45-m radio dish used by \citet{oka98a} and \citet{oka98b} has a beam width
of 16\arcsec~corresponding to $\sim$0.62\,pc and observed the
frequently highly saturated $J=1\rightarrow 0$ $^{12}$C$^{16}$O
emission, the narrow sightline toward the infrared stars (diameters on
the order of 0\farcs00001, $\sim$0.1\,AU), can completely miss
localized dense clouds. Because of the limited sightlines available,
the infrared absorption spectroscopy of H$_3^+$ and CO that we have
employed is not useful for surveys of dense clouds in the CMZ.

Despite its limitations the H$_3^+$ absorption spectrum is a powerful
probe for surveying the warm and diffuse gas which dominates the
CMZ (volume filling factor $\sim$2/3; Paper~I). Although the number of
observed sightlines is limited, each sightline probes a long H$_3^+$
column, as long as the star used as a background source is deeply
embedded in the CMZ. The ($l$, $v$) diagram derived from the H$_3^+$
spectrum provides significant information on the dynamics and morphology
of the gas in the CMZ. Absorption spectroscopy is limited to sampling
gas in front of the stars, but it has the great advantage of
discriminating between motions toward and away from the GC. We note
that most of the H$_3^+$ absorption lines observed toward stars in the GC
are blueshifted and broad, with velocities ranging from $\sim$$-$150 to 
$\sim$$+$10\,km\,s$^{-1}$. This clearly demonstrates that the diffuse gas 
in the CMZ is moving away from the center rather than falling into the center 
as has been suggested for dense gas \citep[e.g.][]{mor96}. Such gas
motions had been noted earlier from radio observations of absorption
by HCO$^+$ toward Sgr~B2 \citep{lin81} and absorption by H$_2$CO
and \ion{H}{1} toward Sgr~A \citep{gus81}. As shown in this paper, 
spectroscopy of H$_3^+$ demonstrates that the expanding gas exists widely in the
CMZ.

The peak optical depths of the strongest lines of H$_3^+$ are at most
0.1. Thus unlike the intense radio CO emission spectrum, where
analysis faces the complication of radiation trapping \citep{sco74, gol74},
 the observed equivalent widths
are linearly related to H$_3^+$ column densities to a good approximation. 
This makes the interpretation of the observed H$_3^+$
spectra direct and simple when viewing dynamical and morphological
data.  The same is true for lines of the $v=2\leftarrow 0$ overtone
band of CO near 2.34\,$\mu$m, unlike the 130 times stronger CO
fundamental band at 4.5--5.0\,$\mu$m and the pure rotational CO
millimeter wave lines.

%------------------------------------------------------------
\subsection{H$_3^+$ as a probe of density and temperature} % subsection 2.6

Although not a key part of this paper and covered in previous papers
(e.g., Paper~I), it is worth summarizing here the unique properties of H$_3^+$
that allow one to unambiguously detect and characterize the low
density and warm temperature of the gas in the CMZ 
(Paper~I). These properties are illustrated by the energy level diagram
in Figure 2 and are described below.
     
At typical cloud temperatures in the Galactic disk (a few tens of
degrees) only the two lowest energy levels of the ground vibrational
state of H$_3^+$ ($J$,$K$) = (1,1) and (1,0) are populated. At higher
temperatures the next three lower levels, (2,1), (2,2), and (3,3) may
come into play. Because H$_3^+$ in the (3,3) level cannot radiatively decay, 
if temperatures are sufficiently high this level has a significant
population and an ro-vibrational absorption line from it at 3.53\,$\mu$m, 
the  $R$(3,3)$^l$ line, is detectable. This situation exists in the CMZ and to 
date has been found nowhere else. On all sightlines observed toward stars 
in the CMZ, the ratio of this line to those from the lowest lying lines from
$J=1$ yield temperatures in the CMZ of $\sim$200 K (Paper~I).

%================================================================
% Figure 2
%================================================================
\begin{figure}
\includegraphics[width=0.48\textwidth, angle=-0]{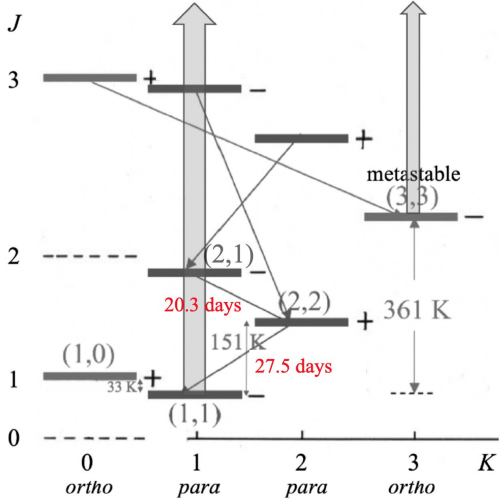}
\caption{Rotational levels of H$_3^+$ in the ground vibrational
  state. This paper is based mainly on velocity profiles of the
  $R$(1,1)$^l$ ro-vibrational transition from the ground (1,1) level (wide arrow),
  with the $R$(3,3)$^l$ line (narrow arrow) providing supplementary
  information. Lifetimes of th (2,1) and (2,2) levels are shown. 
  For the notation $R$(1,1)$^l$ etc. see \citet{lin01}, \citet{oka13}, 
  or \citet{mil20}.}
 
\end{figure}

%================================================================

The (2,1) and (2,2) para-levels lie below the (3,3) level and thus can
be collisionally excited at temperatures at which the (3,3) level is
populated, as well as at somewhat lower temperatures. However, both of
those levels radiatively decay, with lifetimes of $\sim$20 and 27
days, respectively \citep{miz17, nea96}, as illustrated in Figure~2, due
to the spontaneous breakdown of symmetry \citep{oka71, pan86}
The critical densities for these levels are each
$\sim$200\,cm$^{-3}$ \citep{oka04}. At densities lower than that
significant populations in these levels are not maintained and lines
originating from them will be weak or absent. The presence of the
absorption line from (3,3) and the complete absence of lines from
(2,2) on nearly all sightlines in the CMZ (Paper~I) clearly
demonstrate that the CMZ gas in which the H$_3^+$ resides is of low
mean density, most likely $\sim$50\,cm$^{-3}$ (Paper~I).

%============================================================
\section{OBSERVATIONS} % section 3.

Spectra in this series of papers have been obtained using five
infrared spectrographs at five telescopes, three in Hawaii and two in
Chile. Most of the data were acquired by the Phoenix Spectrometer
(resolving power $R\sim$ 60,000) on the 8.1-m Gemini South Telescope. 
Some spectra were obtained by Cold Grating Spectrometer 4 (CGS4, 
$R\sim$ 40,000) on the 3.8-m United Kingdom Infrared Telescope (UKIRT),
the Cryogenic Infrared Echelle Spectrograph (CRIRES, $R\sim$ 50,000 and 
100,000) on the 8.2-m Very Large Telescope (VLT), and the Gemini
Near-Infrared Spectrograph (GNIRS, $R\sim$ 20,000) on the 8.1-m
Frederick C. Gillett Gemini North telescope. Details of the
spectrometers and data reduction are given in Section~2.3. of Paper~I.

%------------------------------------------------------------
\subsection{Bright dust-embedded stars used for the H$_3^+$ spectroscopy} % subsection 3.1

The observations were limited to bright ($L\leq 7.7$\,mag) dust-embedded 
stars with smooth infrared continua in the 3.5--4.0\,$\mu$m
region and bright emission line stars whose lines did not coincide 
with important lines of H$_3^+$. Both types of stars allow sensitive 
high-resolution spectroscopy of the key
lines of H$_3^+$. Altogether spectra of the $R$(1,1)$^l$ line toward
29 stars have been obtained; they are listed in Table~2 of Paper~I. In
this paper, to characterize the gas in front of the central region of
the CMZ from Sgr~A$^\ast$ to 30\,pc to the east, we present spectra of
two stars in the Central Cluster \citep{bec75, bec78, vie05}, 
two stars between the Central
Cluster and the Quintuplet Cluster, and five stars in the Quintuplet
Cluster \citep{kob83,  nag90, oku90}. Three of these are NHS stars, 
found by \citet{nag93} and one is an FMM star \citep{fig99}. 
All nine are high mass stars and all but two were known prior
to the \citet{geb19a} survey. All exhibit strong absorption
lines of H$_3^+$ indicating simultaneously their similar depths in the
CMZ, the ubiquity of H$_3^+$ in the gas in front of the central
30\,pc, and the abundance of the warm and diffuse gas in which that
H$_3^+$ resides.

To study the gas in the rest of the CMZ we have observed 13 of the
suitable stars scattered across the CMZ, from 140\,pc to the
west of Sgr~A$^\ast$ to 116\,pc to the east, newly found by the \citet{geb19a} 
survey. The spectra of nine of them are presented
here. Several additional stars were found to be foreground stars, but
based on the strengths of their H$_3^+$ absorption lines, these nine
stars are located deeply enough in the CMZ to provide valuable
data. All of the newly found stars were known only by their
coordinates in the Two-Micron All Sky Survey (2MASS) catalogue. For
convenience of memory and discussion in Paper~I  we unofficially
designated the newly found stars using the Greek alphabet $\alpha$ to
$\lambda$ from the west to east. As the survey found more suitable
stars we designated them by $\alpha-$, $\alpha+$, $\alpha+-$, etc. 
We use these designations here.

The newly found stars tend to be clustered. For example, close to the
western edge of the CMZ and nearly on the Galactic plane three
suitable stars were found ($\alpha$, $\alpha+$, and $\beta$).  No
stars were found that close to the eastern edge; the closest are stars
$\lambda-$ and $\lambda-+$. The stars do not provide an ideally
sampled ($l$, $v$) diagram, but still provide enough information for a
detailed study to be feasible and for key conclusions to be drawn.

The 18 stars whose spectra are reported here are listed in
Table~1. The table includes the $L$ magnitudes of the stars, their
Galactic longitudes, their distances from Sgr~A$^\ast$ on the plane of
the sky in the longitudinal (EW) direction, their latitudinal (NS)
distances from the Galactic plane, and the ranges of LSR absorption
velocities of warm diffuse gas observed in H$_3^+$.

%============================================================
% Table 1
%============================================================
\begin{deluxetable*}{lcrrrll}
\tablecaption{Observed H$_3^+$ $R$(1,1)$^l$ Spectra \label{t1}}
% \tableline \tableline
\tablehead{
  \colhead{Stars                                         } &
  \colhead{$L$ (mag)\tablenotemark{a}                    } &
  \colhead{$G_{\rm lon}$ (\degr)                         } &
  \colhead{EW (pc)                                       } &
  \colhead{NS (pc)                                       } &
  \colhead{Velocity range (km\,s$^{-1}$)\tablenotemark{b}} &
  \colhead{Note}
}
\startdata
% \multicolumn{4}{c}{H$_3^+$} \\
% \cline{1-4}
% Star         & $L$ (mag)a  &  $G_{\rm lon}$ (\degr) &  EW (pc)  & NS (pc)   & Velocity range (km\,s$^{-1}$)b    & Note         \\
$\alpha$     & 3.79        & $-$1.0463              & $-$138.7  &  $-$2.65  & ($-$15,   15)                      & narrow line  \\ % α       
$\alpha+$    & 7.02        & $-$1.0417              & $-$138.0  & $-$6.12   & ($-$25,  25)                      & narrow line  \\ % α+       
$\beta$      & 4.53        & $-$1.0082              & $-$133.4  & $-$4.20   & ($-$15,   15)                      & narrow line  \\ % β        
$\gamma$     & 6.41        & $-$0.5442              & $-$68.4   & $-$6.48   & ($-$165, $-$110)                  & narrow line  \\ % γ        
$\delta$     & 6.51        & $-$0.2834              & $-$31.9   & 1.57      & ($-$163, $-$116) ($-$93, 26)      & two troughs  \\ % δ         
GCIRS\,3     & 4.84        & $-$0.0576              & $-$0.15   & 0.07      & ($-$150, $\sim$0) & trough  \\ % GCIRS 3  
GCIRS\,1W    & 4.92        & $-$0.0549              & 0.06      & 0.15  & ($-$145, $\sim$0)  & trough  \\ % GCIRS 1W 
$\epsilon+$  & 7.56        & 0.0300                 & 12.0      & $-$5.88   & ($-$158, $\sim$0)                 & trough       \\ % ε+        
NHS\,21      & 4.60        & 0.0992                 & 21.7      & $-$1.27   & ($-$145, $-$103) ($-$90, $\sim$0) & two troughs  \\ % NHS 21   
NHS\,42      & 6.61        & 0.1479                 & 28.5      & 0.50      & ($-$147, 17)                      & trough       \\ % NHS 42   
NHS\,25      & 5.63        & 0.1591                 & 30.1      & $-$2.69   & ($-$156, 16)                      & trough       \\ % NHS 25   
$\eta$       & 5.52        & 0.1631                 & 30.6      & $-$3.92   & ($-$160, $\sim$0)                 & trough       \\ % η         
GCS\,3-2     & 3.16        & 0.1635                 & 30.7      & $-$2.00   & ($-$145, 15)                      & trough       \\ % GCS 3-2  
FMM\,362     & 6.41        & 0.1787                 & 32.8      & $-$2.67   & ($-$145, $-$20)                   & trough       \\ % FMM 362  
$\theta$     & 6.38        & 0.2647                 & 44.9      & $-$3.86   & ($-$160, $-$30)  ($-$15, 15)      & two troughs   \\ % θ         
$\iota$      & 6.58        & 0.5477                 & 84.5      & $-$1.83   & ($-$115, $-$40)                   & trough        \\ % ι         
$\lambda-$   & 6.72        & 0.7685                 & 115.5     & $-$28.54  & ($-$52,     20)                   & trough       \\ % λ−        
$\lambda-+$  & 7.06        & 0.7746                 & 116.2     & 21.98     & ($-$80,  $\sim$0)                 & trough       \\ % λ−+       
\enddata
\tablenotetext{a}{Magnitudes from GLIMPSE IRAC\,1 filter, Ramirez et al. 2008), except from Viehmann et al. (2005) for GCIRS\,1W and GCIRS\,3.}
\tablenotetext{b}{Values within parentheses are in LSR and are minima and maxima of continuous ranges of absorption.}
\end{deluxetable*}

%============================================================

%------------------------------------------------------------
\subsection{Observed velocity profiles of H$_3^+$ : the CMZ's contribution}
% subsection 3.2

In this subsection we illustrate the general approach to understanding
the absorption spectra of H$_3^+$ lines toward stars in the CMZ by
presenting and analyzing the velocity profiles of H$_3^+$ and CO
absorption lines toward the bright infrared source GCS\,3-2, located in
the Quintuplet Cluster. Profiles of some of the lines observed toward it 
are shown in Figure~3.

%============================================================
% Figure 3
%============================================================
\begin{figure}
\includegraphics[width=0.50\textwidth]{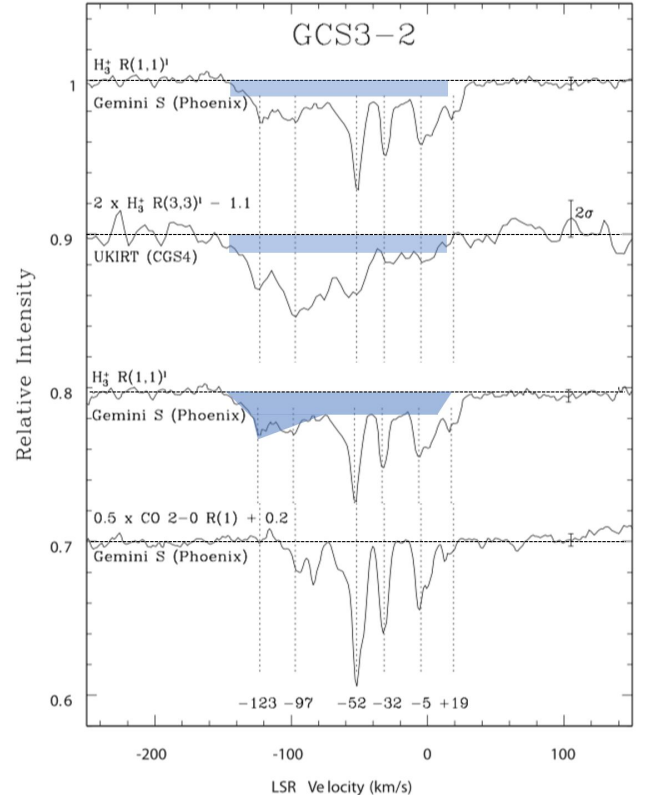}
\caption{Velocity profiles of H$_3^+$ and CO lines toward the star
  GCS\,3-2 adaped from \citet{oka05}). Horizontal dashed lines indicate the
  continuum. Top trace: $R$(1,1)$^l$ profile (see
  Section~3.2.1.). Second trace: $R$(3,3)$^l$ profile (see
  Section~3.2.2.). Third trace: same as top trace, but with
  absorptions of H$_3^+$ in the dense clouds in the foreground spiral
  arms subtracted, to produce the shaded area, which is that part
  of the $R$(1,1)$^l$ profile originating in the CMZ. Bottom trace: 
  CO $v=2\leftarrow0$ $R$(1) profile, with
  narrow and deep absorptions produced mainly at approximately $-$52,
  $-$32, and 0\,km\,s$^{-1}$ by dense clouds in the three foreground
  spiral arms. The shaded rectangles in the upper two spectra depict 
  the velocity extent of the warm diffuse CMZ gas.}
\end{figure}
%============================================================

%------------------------------------------------------------
\subsubsection{The $R$(1,1)$^l$ spectrum: subtraction of spectra in spiral arms}
% subsubsec 3.2.1

For investigating velocity profiles of spectral lines of H$_3^+$, the
$R$(1,1)$^l$ transition at 2691.443\,cm$^{-1}$ (3.71548\,$\mu$m, thick
arrow in Figure~2) arising from the ground (1,1) level and having the
transition dipole moment 0.119\,D, is the optimal choice; it is the
strongest line, it is relatively free from atmospheric interference,
and its intensity is nearly independent of temperature.

A complication in interpreting the absorption velocity profile of this line
is that absorption by H$_3^+$ can be produced not only in the CMZ, but
also in the three foreground spiral arms (see the top trace of
Figure~3). Fortunately, in dense clouds absorption lines are also
produced by CO, whose overtone band can be used to clearly identify
the contribution of foreground absorption to the H$_3^+$ line
profiles. In addition the absorption lines in spiral arm clouds are
almost always narrow and occur at characteristic radial velocities. 
These aspects are illustrated in Figure~3, where
one can see three sharp optically thin absorption components in a
2.34-$\mu$m CO overtone line (bottom trace), and similar features
contributing to the H$_3^+$ $R$(1,1)$^l$ line profile (top trace). The
three sharp absorptions in both spectra at $\sim$ $-$50\,km\,s$^{-1}$,
$\sim$ $-$30\,km\,s$^{-1}$, and $\sim$0\,km\,s$^{-1}$ correspond to
the velocities of gas in the 3\,kpc arm \citep{rou60}, the
4.5\,kpc arm \citep{men70} and the local arm,
respectively. Each of these absorptions has a small velocity gradient
with longitude, which is used for studies of kinematics and dynamics
of the gas \citep[e.g.][]{sof06}. It is relatively straightforward to
subtract them from the H$_3^+$ $R$(1,1)$^l$ profile to elucidate the
spectrum of H$_3^+$ in the CMZ. After the subtraction, the $R$(1,1)$^l$
absorption is an uneven trough with the approximate shape of the 
shaded polygon in the third trace of Figure~3.

As will be illustrated in Section 4, on sightlines more distant from
the central region, the velocity dispersion of the CMZ gas decreases
with distance. Near the western edge of the CMZ, where suitable stars 
are available to probe the gas, the dispersion nearly vanishes, as the
motion of the gas is largely perpendicular to the line of sight. At
these locations it is more challenging to separate the absorption in the
$R$(1,1)$^l$ line due to CMZ gas from that of local arm (except for
the case of the sightline toward Star $\alpha$; see Section~4.7). In
these cases observations of the $R$(3,3)$^l$ line, which is absent in
the low temperature gas of the spiral arms, are critical for
identifying the presence of low velocity warm diffuse gas in the CMZ.
More details on the spectral features arising in the foreground spiral 
arms can be found in Section~5.3.1 of Paper~I.

%------------------------------------------------------------
\subsubsection{The $R$(3,3)$^l$ spectrum: the fingerprint of H$_3^+$in the CMZ} 
%subsubsec %3.2.2
   
The $R$(3,3)$^l$ line at 2829.935\,cm$^{-1}$ (3.53365\,$\mu$m, narrow
arrow in Figure~2), with transition dipole moment 0.138\,D, arises from
the metastable (3,3) level of ortho-H$_3^+$. It is generally weaker
than the $R$(1,1)$^l$ line but its profile is free from contamination by
foreground absorption because of the low temperature of gas in the
spiral arms. Thus, it is the fingerprint by which H$_3^+$ in the 
warm gas of the CMZ can be unambiguously identified.

Observing this line from ground-based telescopes can be challenging,
however. The frequency of the $R$(3,3)$^l$ line closely coincides with
the strong $2\nu_{2}\,5_{05} \leftarrow 6_{34}$ transition of H$_2$O at
2830.008\,cm$^{-1}$. The difference of 0.073\,cm$^{-1}$ corresponds to
a Doppler shift of only 8\,km\,s$^{-1}$. Since the distribution of
atmospheric H$_2$O varies temporarily depending on the weather, and
because the telluric absorption is strong and broad, correction for
this line is more difficult than correcting for telluric absorption
lines of other molecular species, especially at low radial velocities. 
This usually makes observations of
the $R$(3,3)$^l$ line from telescopes at the 4200-m summit of Maunakea 
Hawai`i (Subaru, Gemini North, UKIRT) more dependable than those from 
Cerro Pachon (2715\,m Gemini South) and Cerro Paranal (2635\,m VLT) Chile.

The intensity of the $R$(3,3)$^l$ line relative to that of the $R$(1,1)$^l$ line is 
sensitive to the temperature of the gas. In the top trace of Figure~3 note
that while the strength of the $R$(1,1)$^l$ absorption trough toward
GCS\,3-2  (as well as those of several other centrally located 
stars) is roughly constant across most of its observed velocity range, the depth 
of the $R$(3,3)$^l$ line (second trace in Figure 3) first increases 
toward more negative radial velocities until $\sim-100$ km s$^{-1}$.
This  indicates increasing temperature of the gas 
with more negative velocities over the above range.
%============================================================
\section{THE SPECTRA} % section 4

In this section we present and describe the H$_3^+$ $R$(1,1)$^l$ line
profiles toward the 17 of the 18 selected stars (for GCS\,3-2 see the
preceding section). We begin with the two stars in the Central
Cluster, move east to two stars in the region between the Central
Cluster and the Quintuplet Cluster, and then further east to four
stars in the Quintuplet Cluster itself. We then examine the
$R$(1,1)$^l$ velocity profiles of two stars approximately midway
between the center and the eastern edge of the CMZ, and finally those
of the two stars closest to the eastern edge, although still some
25\,pc distant from the edge. To the west of the Central Cluster we
show the $R$(1,1)$^l$ profiles of two stars located between the center
and the western edge of the CMZ, and lastly, the profiles of three
stars located very close to the western edge.

In determining the presence and extent of the warm diffuse gas, in 
almost all cases we rely on the widths of the absorption troughs  of the 
$R$(1,1)$^l$ lines presented here and the published profiles of the 
$R$(3,3)$^l$ line and/or CO overtone lines for guidance. We do not 
present the profiles of the $R$(3,3)$^l$ line and CO lines here; the 
reader is referred to spectra of those lines  
in \citet{got08}, \citet{geb10}, \citet{got11}, \citet{got14}, and Paper I.

%------------------------------------------------------------
\subsection{Central Cluster : GCIRS\,3 and GCIRS\,1W} % subsection 4.1

Surrounding the black hole Sgr~A$^\ast$ and within a radius on the sky
of 0.4\,pc are many bright stars of the Central Cluster \citep[see Fig.~2 of][]
{vie05}, spectra of two of which are shown in Figure~4,
GCIRS\,3 (upper trace) and GCIRS\,1W (lower trace). A significant
difference between the sightlines toward these stars and that toward
GCS\,3-2 is that while the latter probes only diffuse clouds, the
former sightlines cross dense gas in the circumnuclear disk \citep[CND;][]{gen87}, 
which produces the absorption features in the
two lines shown as well as in the $R$(2,2)$^l$ line, at
$\sim$50\,km\,s$^{-1}$ in GCIRS\,3 and $\sim$40\,km\,s$^{-1}$ in
GCIRS\,1W \citep{got14}. Discussions of these features is outside
the scope of this paper.

%============================================================
% Figure 4
%============================================================
\begin{figure}
\includegraphics[width=0.48\textwidth]{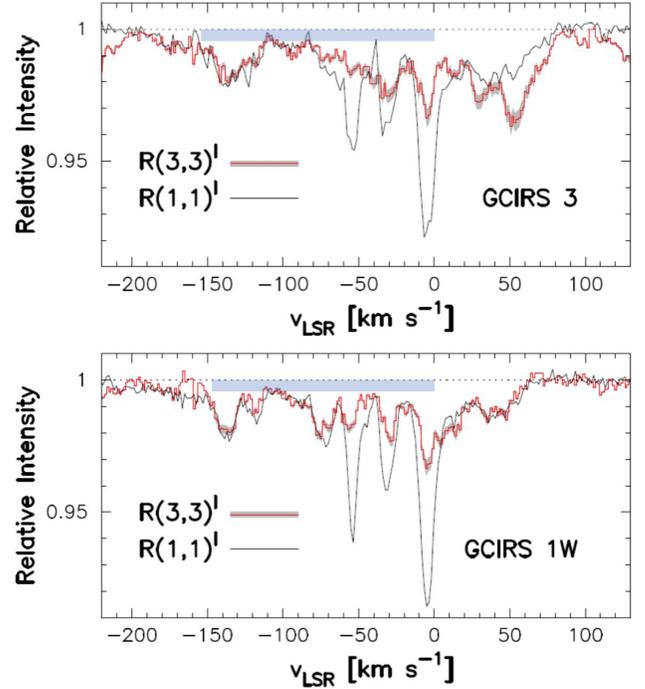}
\caption{Velocity profiles of the $R$(1,1)$^l$ and $R$(3,3)$^l$ lines
toward the Central Cluster stars GCIRS\,3 (top) and GCIRS\,1W (bottom),
from \citet{got14}). The  $R$(3,3)$^l$ absorption  has been scaled by a factor
of 1.3 for ease of compairison. All spectra were recorded by the CRIRES
spectrometer at the VLT. Shaded rectangles indicate the extents 
of the warm diffuse CMZ gas (see Section 4.1).}
\end{figure}
%============================================================

Like the spectra of H$_3^+$ lines toward GCS\,3-2, spectra of GCIRS\,3
and GCIRS\,1W exhibit broad absorption troughs due to warm diffuse gas 
extending from $-$150\,km\,s$^{-1}$ to near 0\,km\,s$^{-1}$. At positive velocities the
troughs merge with the absorption by the dense CND gas mentioned above. Therefore the
velocities at these edges of the troughs are uncertain,
and could be slightly positive. Also, unlike the spectra toward
GCS\,3-2 where the  depth of the absorption trough varies fairly smoothly with
velocity, the profiles toward GCIRS\,3 and GCIRS\,1W show
significant variations of depth, indicating variations in the density
of the warm diffuse gas with velocity, and presumably with radial
location within the CMZ on these sightlines. Interestingly both the
$R$(1,1)$^l$ and $R$(3,3)$^l$ lines have absorption features near the
radial velocities of the three foreground spiral arms. On the other
hand, spectra of CO overtone lines toward these sources also have prominent 
absorptions at all three velocities ($-$50, $-$30, and 0 \,km\,s$^{-1}$), 
while spectra of the $R$(2,2)$^l$ line show no absorption except at the 
positive velocity of the CND \citep{got14}. We
therefore conclude that much of the deep absorptions in the
$R$(1,1)$^l$ line profile at these velocities is produced by dense gas in the spiral
arms. However, because the metastable (3,3) level is not populated
in the cold gas of the spiral arms, we also conclude that these three
absorption features in the $R$(3,3)$^l$ line  are formed in the warm
and diffuse gas in the CMZ, and only coincidentally approximately
match the velocities of the foreground spiral arms.

%============================================================
% Figure 5
%============================================================
\begin{figure}
\includegraphics[width=0.47\textwidth]{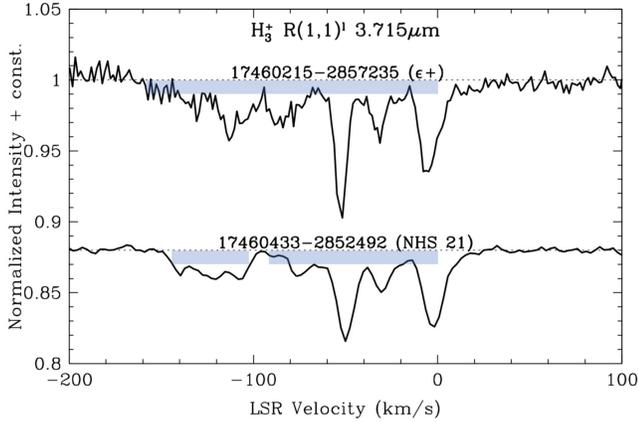}
\caption{The velocity profiles of the $R$(1,1)$^l$ line of H$_3^+$
observed toward the Stars $\epsilon+$ and NHS\,21. Both spectra were
obtained by the Phoenix Spectrometer at the Gemini South Observatory.
Shaded rectangles indicate the extents of the warm diffuse CMZ gas (see Section 4.2).}
\end{figure}

%============================================================

%------------------------------------------------------------
\subsection{Stars between the Central Cluster and the Quintuplet Cluster : $\epsilon+$ and NHS\,21}
% subsection 4.2
    
Figure~5 contains the velocity profiles of the $R$(1,1)$^l$ spectra
toward two stars located between the Central Cluster and the
Quintuplet Cluster, $\epsilon+$ and
NHS\,21. Absorption from the former contains a broad shallow feature
similar to that toward GCS\,3-2, as well as the three sharp spiral arm
features; the latter are also present in the spectrum of NHS\,21. The
broad absorption in the spectrum of NHS 21, however, has a
gap from $–$103\,km\,s$^{-1}$ to
$-$90\,km\,s$^{-1}$, with little or no absorption. It is tempting to
conclude from this that NHS\,21 is more shallowly embedded in the
CMZ, but the similar depths of the troughs suggest otherwise. We
suspect that NHS\,21 lies deep within the CMZ, but that warm diffuse gas
in the above narrow velocity range is simply absent on sightlines though that region of the CMZ.

%============================================================
% Figure 6
%============================================================
\begin{figure}
\includegraphics[width=0.48\textwidth]{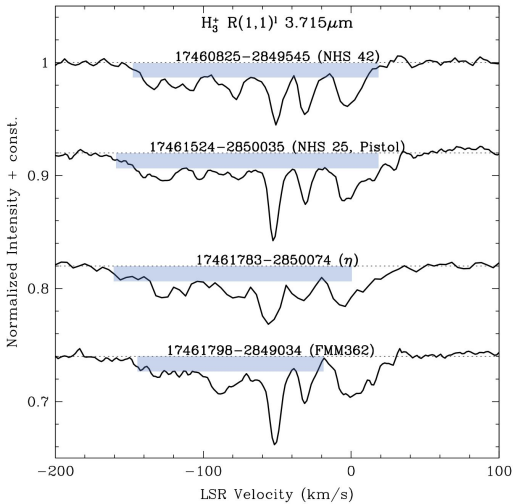}
\caption{Velocity profiles of the $R$(1,1)$^l$ lines observed toward
four stars in the Quintuplet Cluster. All of the spectra were recorded
by the Phoenix Spectrometer on the Gemini South telescope.
Shaded rectangles indicate the extents of the warm diffuse CMZ gas (see Section 4.3).}
\end{figure}

%============================================================

%------------------------------------------------------------

\subsection{Quintuplet Cluster : GCS\,3-2, NHS\,42, NHS\,25, $\eta$, FMM\,362}
% subsection 4.3

Spectra of H$_3^+$ lines toward four of the five bright and closely
spaced infrared Quintuplet stars, for which the Cluster is named:
GCS\,3-1, GCS\,3-2, GCS\,3-4, and GCS\,4, were published in
Paper~I. The four are within 0.7\,pc of one another on the plane 
of the sky. Their velocity profiles are
nearly identical (Paper~I, Figure~3); Figure~3 of this paper, showing
spectra of GCS\,3-2, is a representative example. As discussed in
Section~3.2, and illustrated in Figure~3, the CMZ produces a broad and
almost entirely blueshifted absorption toward GCS\,3-2 and the other
members of the Quintuplet.

Figure~6 contains the velocity profiles of the H$_3^+$ $R$(1,1)$^l$
line toward four other stars within the Quintuplet Cluster. Two of the
stars, NHS\,42 and FMM\,362, are situated near opposite edges of the
Cluster, which has a diameter of $\sim$2\,pc (Figer, McLean, \& Morris
1999). Once the foreground sharp spiral arm absorptions are removed, 
each of the four reveals a broad absorption trough of extent and
shape generally similar to that of GCS\,3-2. Many of the bright Quintuplet
stars are known to possess energetic winds. However, the overall
similarity of the CMZ profiles both toward them and toward the stars
presented in Section~4.2, which are as much as 20\,pc distant in the
plane of the sky, is a convincing demonstration that the absorptions
are not due to localized gas motions associated with any of the stars
but are part of a much larger scale phenomenon.

%============================================================
% Figure 7
%============================================================
\begin{figure}
\includegraphics[width=0.5\textwidth]{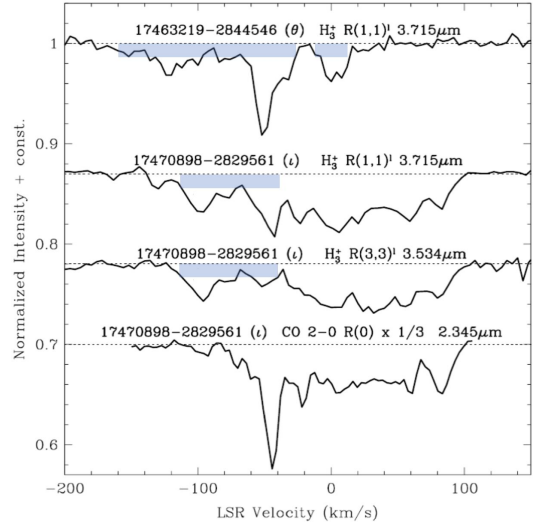}
\caption{The velocity profiles of the $R$(1,1)$^l$ line observed
toward the Star $\theta$ (top trace), and of the $R$(1,1)$^l$ (second
trace), $R$(3,3)$^l$ (third trace), and CO $v=2\leftarrow 0$ $R$(0)
(bottom trace) lines observed toward Star $\iota$ near Sgr~B. All spectra 
were recorded by the Phoenix Spectrometer at the Gemini South Observatory.
Shaded rectangles indicate the extents of the warm diffuse CMZ gas 
(see Section 4.4).}
\end{figure}
%============================================================

%------------------------------------------------------------
\subsection{Stars roughly halfway to the eastern edge of the CMZ : $\theta$ and $\iota$}
%subsection 4.4

Figure~7 contains spectra toward two stars located 45\,pc and 85\,pc
to the east of Sgr~A$^\ast$. They lie between the Quintuplet Cluster
and the suitable stars found in the survey that are closest to
the eastern edge of the CMZ. Very roughly speaking these two stars are
halfway between Sgr~A$^\ast$ and the eastern edge of the CMZ.

Interpretation of the $R$(1,1)$^l$ spectrum toward Star $\theta$ (top trace), 
which is the nearer of the two to the center, is somewhat of a challenge. Blueshifted absorption 
in this line extends to $\sim$ $-$160\,km\,s$^{-1}$. Its assignment to warm 
diffuse gas is consistent with our (unpublished) spectrum of the $R$(3,3$)^l$ line, which
shows absorption to approximately that velocity.  At velocities more positive than
$-$60\,km\,s$^{-1}$ absorption is contributed by H$_3^+$ in the spiral arms near 
$-$51\,km\,s$^{-1}$ (3\,kpc arm), $-$32\,km\,s$^{-1}$ (4.5\,kpc arm)
and $-$1\,km\,s$^{-1}$ (local arm), as attested to by the sharp absorption features in the 
CO overtone lines (see Fig. 6 of Paper I).  However, the deep shoulder between the
$-$51\,km\,s$^{-1}$ and $-$32\,km\,s$^{-1}$ features, which is not
observed toward other stars, suggests that the trough of absorption by warm diffuse gas 
extends to $\sim$ $-$30\,km\,s$^{-1}$. The $R$(3,3$)^l$ line profile also 
contains absorption in the $-$60 to $-$30 \,km\,s$^{-1}$ interval. In addition, both the
$R$(1,1$)^l$ and $R$(3,3$)^l$ lines show absorption from  $-$15 to $+$15 \,km\,s$^{-1}$.
The lack of absorption near $-$20\,km\,s$^{-1}$ implies that the trough has a narrow 
gap centered near that velocity.

The sightline toward Star $\iota$ (2MASS\,17470898$-$2829561), located
midway between Sgr~B1 and Sgr~B2, is unique in showing very broad, deep, 
and structured absorption profiles in lines of both H$_3^+$ and the CO overtone 
band, extending from $\sim$ $-$100\,km\,s$^{-1}$  to $\sim$ $+$100\,km\,s$^{-1}$ 
\citep[lower three traces of Figure 7; see also][]{geb10}. The strength of the 
CO absorption indicates that the CO is located in a number of dense clouds on
this sightline. Because at positive velocities the pure rotational CO $J$ = $1-0$ 
emission profile in Sgr B at this location \citep{oka98a} roughly resembles these 
infrared profiles, Star $\iota$ must be located within the Sgr B cloud complex. 
The sharp absorption at $-$43\,km\,s$^{-1}$ in the CO $R$(0) line, other low $J$ 
CO lines (Fig. 6 of Paper I), and in the 
H$_3^+$ $R$(1,1) line together with its absence in the H$_3^+$ $R$(3,3) line demonstrate that 
this feature is produced by dense gas in the 3 kpc arm.  Nearly all of the rest of the 
absorption in this nearly 200 \,km\,s$^{-1}$ wide interval arises within the CMZ. 

At  the  highest negative velocities, between $-70$ and $-115$\,km\,s$^{-1}$, absorption in 
the  $R$(1,1)$^l$ and $R$(3,3)$^l$ lines is strong, but is virtually absent 
in CO (Figure 7) and in the $R$(2,2)$^l$ line (Paper I, Fig. 8). Thus the absorbing H$_3^+$ 
in this velocity interval is located in warm and diffuse gas in the CMZ. Absorption in the 
$R$(2,2)$^l$ line is also weak or absent from $-70$ to $-$40\,km\,s$^{-1}$ (Paper I). 
Therefore, the $R$(3,3)$^l$  absorption from $-$70\,km\,s$^{-1}$ to $-$40\,km\,s$^{-1}$ 
is also produced largely in diffuse gas. Thus, as found elsewhere in the CMZ, 
warm diffuse gas is present on this sightline over a wide range of negative velocities, in 
this case definitely from $-115$ \,km\,s$^{-1}$ to $-40$ \,km\,s$^{-1}$. Warm diffuse 
gas may also be present at less negative  than $-$40\,km\,s$^{-1}$ as  it is elsewhere; 
however, its unique signatures are hidden by the signatures of 
warm dense gas in this velocity range. Additional spectra and discussion of this 
interesting sightline will be presented in a separate paper (T. R. Geballe et al. in preparation).

%------------------------------------------------------------
\subsection{Stars Nearest to the Eastern Edge : $\lambda-$ and $\lambda-+$}

The two stars whose $R$(1,1)$^l$ spectra are shown in Figure~8,
$\lambda-$ and $\lambda-+$, are each $\sim$25\,pc from the eastern
edge of the CMZ. They are also much further displaced (in latitude)
from the Galactic plane than any of the other stars in the sample. Nevertheless, 
their spectra provide evidence that the sightline to each of them passes through CMZ gas.

The $R$(1,1)$^l$ line profile toward Star $\lambda-$ (upper trace) shows absorption 
between $\sim$$-$90\,km\,s$^{-1}$ and $\sim$$+$20\,km\,s$^{-1}$.
The detection of absorption between $\sim$$-$90\,km\,s$^{-1}$ and $\sim$$-$60\,km\,s$^{-1}$
is clearly marginal; however the profile of the $R$(3,3)$^l$ line in Paper I (Fig. 7) also shows 
continuous absorption in this velocity range (and extending continuously into positive velocities), 
and we therefore conclude that warm diffuse gas is present to velocities as far negative as 
$-$90\,km\,s$^{-1}$. Absorption by dense gas in the foreground spiral arms may be present, 
but the broad $R$(3,3)$^l$ line profile clearly demonstrates the existence of warm gas over its wide 
range of absorption velocities.

%============================================================
% Figure 8
%============================================================
% The velocity profiles => Velocity profiles?
\begin{figure}
\includegraphics[width=0.5\textwidth]{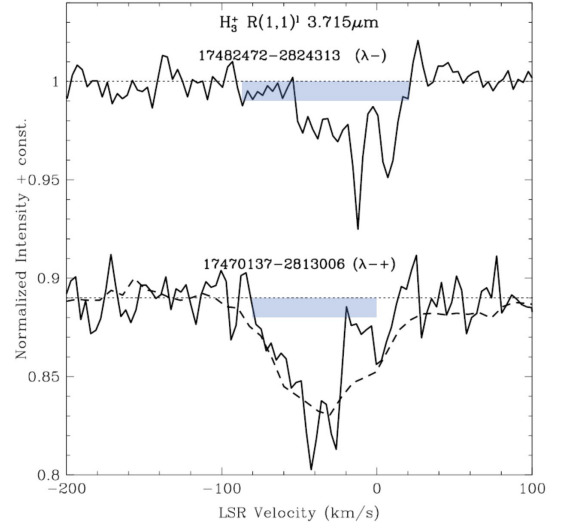}
\caption{The velocity profiles of the $R$(1,1)$^l$ line observed
toward the $\lambda-$ and $\lambda-+$. Both spectra were recorded by
the Phoenix Spectrometer at the Gemini South Observatory. The low
resolution dotted spectrum in the lower trace was recorded by the
GNIRS spectrometer at the Gemini South Observatory.
Shaded rectangles indicate the extents of the warm diffuse CMZ gas (see Section 4.5).}
\end{figure}
%============================================================

The higher-resolution and lower-resolution $R$(1,1)$^l$ spectra toward Star
$\lambda-+$ (lower trace) are generally consistent with one
another. Both spectra reveal an absorption trough starting at $\sim$
$-$80\,km\,s$^{-1}$.  Although it is likely that there is
contamination of the profile by foreground dense spiral arm gas at
$\sim$ $-$45\,km\,s$^{-1}$ and $-$25\,km\,s$^{-1}$, the trough may
extend to 0\,km\,s$^{-1}$.

Thus, the velocity extents of the troughs toward both of these stars
are somewhat uncertain. However, the key conclusion from these spectra is that
the absorptions by warm diffuse CMZ gas in these easternmost stars do not
extend to the highest negative velocities seen toward each of the other
eastern stars closer to the Central Cluster.

%------------------------------------------------------------
\subsection{Stars Between the Central Cluster and the Western Edge of the CMZ : 
$\delta$ and $\gamma$}
% subsection 4.6.
  
We now consider the spectra of stars to the west of the Central
Cluster that surrounds Sgr~A$^\ast$. Compared to the eastern side of
the CMZ, fewer suitable stars are available \citet{geb19a}. Several of
them were found to exhibit little
or no absorption due to H$_3^+$, indicating that they are not deeply
embedded in the CMZ (see footnote for Table 2 Paper~I).  However, the
spectra of H$_3^+$ toward two stars located between the center and the
western edge and three stars that are very close to the western edge
provide important information on the extent and kinematics of the warm
diffuse gas in the CMZ.

%============================================================
% Figure 9 
%============================================================
\begin{figure}
\includegraphics[width=0.5\textwidth]{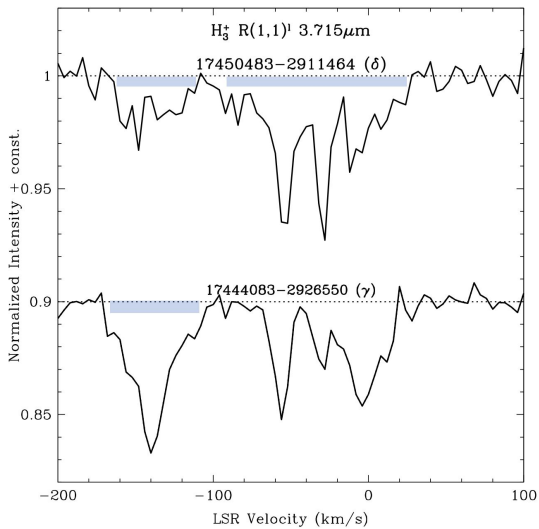}
\caption{The velocity profiles of the $R$(1,1)$^l$ line observed
toward the stars $\gamma$ and $\delta$. Both spectra were recorded by
the Phoenix Spectrometer in the Gemini South Observatory.
Shaded rectangles indicate the extents of the warm diffuse CMZ gas
(see Section 4.6).}
\end{figure}
%============================================================

Figure~9 contains the spectra of the $R$(1,1)$^l$ line towards the two
intermediate western stars. Star $\delta$ (upper trace) is located
some 30\,pc from Sgr~A$^\ast$, at the same distance from it as is the
Quintuplet Cluster on the opposite side. Like the stars in that cluster,
its spectrum contains an extensive absorption trough, from
$\sim$$-$160\,km\,s$^{-1}$ to $\sim$20\,km\,s$^{-1}$, albeit with
 narrow gap in the absorption from $-$110\,km\,s$^{-1}$ to $\sim$
$-$90\,km\,s$^{-1}$.

The $R$(1,1)$^l$ velocity profile toward Star $\gamma$ (Figure~9, 
lower trace), in the Sgr~C complex located $\sim$70\,pc to the west of 
Sgr~A$^\ast$ and thus approximately midway between the center and the 
western edge of the CMZ, is unique among all of the observed spectra in
that it is entirely highly blueshifted, extending from  $-$165\,km\,s$^{-1}$ to 
$-$110\,km\,s$^{-1}$, with no trough extending to less negative velocities. 
The same broad absorption is also present in the 
spectrum of the $R$(3,3)$^l$ line (Paper I, Fig. 7). The absorption in the $R$(1,1)$^l$ lines
has a sharp peak centered  at $-$136\,km\,s$^{-1}$.
The triangular profile of the $R$(1,1)$^l$ absorption line is reminiscent of 
the OH radio absorption line in Fig.~2 of \citet{bol64b}. Although the
CO spectra on this sightline show narrow absorption lines centered at 
at $-$136\,km\,s$^{-1}$, they are rather weak; moreover there is no absorption in the 
$R$(2,2)$^l$ line of H$_3^+$ (see Figures~6 and 8 and Table~3 of
Paper~I). Therefore, although the sharp $R$(1,1)$^l$ absorption peak at  
 $-$136\,km\,s$^{-1}$ may be partly due to the contribution of a dense cloud, the bulk of 
the absorption profile must be due to H$_3^+$ in the diffuse gas of the CMZ. The lack of an
absorption trough extending from $-$90\,km\,s$^{-1}$ to less negative
velocities is discussed further in Section~5.

%============================================================
% Figure 10
%============================================================
\begin{figure}
\includegraphics[width=0.5\textwidth, angle=-0]{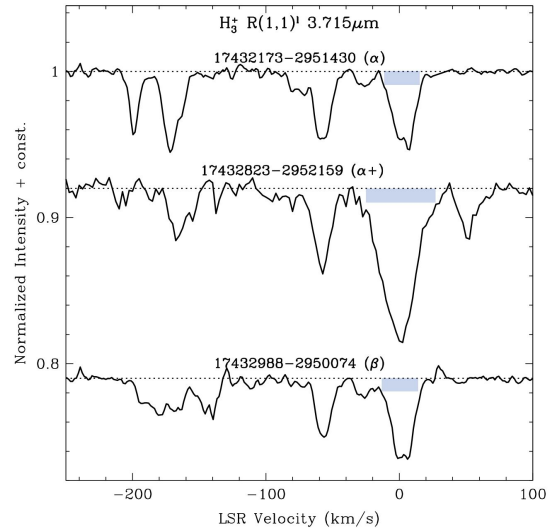}
\caption{Velocity profiles of the $R$(1,1)$^l$ line observed toward
three stars at the western edge of the CMZ.  All spectra were recorded
by the Phoenix Spectrometer at the Gemini South Observatory.
Shaded rectangles indicate the extents of the warm diffuse CMZ gas
(see Section 4.7).}
\end{figure}
%============================================================

%------------------------------------------------------------
\subsection{Stars at the Western Edge of the CMZ : $\alpha$, $\alpha+$, and $\beta$}
% subsection 4.7

Velocity profiles of the $R$(1,1)$^l$ line of H$_3^+$ toward three
bright stars, $\alpha$, $\alpha+$, and $\beta$ at the western edge of
the CMZ and close to the radio complex Sgr~E are shown in
Figure~10. They, together with spectra of the $R$(3,3)$^l$ line and 
overtone CO lines toward them presented in \citet{geb10}, \citet{got11}, and 
in Paper~I, have played an essential role in our overall
understanding of dynamics and morphology of the CMZ warm and diffuse
gas.

None of the spectra of Stars $\alpha$, $\alpha+$, and $\beta$ contains
the broad absorption troughs that are so characteristic of almost all
of the spectra toward stars that are not on the edges of the
CMZ. Instead each of these stars contains three fairly narrow
absorption features at negative velocities, and a strong absorption
centered at 0\,km\,s$^{-1}$. This last absorption component, highly
important in understanding the kinematics of the warm and diffuse gas
throughout the CMZ, is discussed following descriptions of the
other absorption features.

%------------------------------------------------------------
\subsubsection{Absorption by localized dense clouds in the CMZ}
% subsubsec 4.7.1.
  
The two most negative absorption velocities, at $\sim$
$-$205\,km\,s$^{-1}$ and $\sim$ $-$170\,km\,s$^{-1}$ (with some
variation from one sightline to another) are seen toward all three
stars. They match the velocities of two high velocity dense clouds
previously observed by Liszt (1992) in $^{13}$CO (see the ($l$, $v$)
diagram of his Fig.~3). They also match absorption velocities in
spectra of overtone CO lines toward some of these sources (Geballe \&
Oka 2010; Paper~I). In addition, emission at these velocities is also
seen in far-infrared emission lines of \ion{C}{2} and \ion{N}{2} in
this region \citep{lan15}, which must arise in diffuse gas
associated with the dense clouds. A third strong negative velocity
absorption in the H$_3^+$ spectra, near $-$60\,km\,s$^{-1}$, is due to
foreground dense gas in the 3\,kpc spiral arm. All of these dense
clouds are of minor interest for this paper and are not discussed
further. In addition to them, the $R$(1,1)$^l$ spectrum toward Star
$\alpha+$ (Figure~10, middle trace) contains a narrow absorption
centered at $+$51\,km\,s$^{-1}$, not present in the spectra of the
other two stars.

All of the absorptions at negative velocities are strongly present in 
spectra of the overtone CO lines (\citet{oka19} and unpublished data),
further demonstrating that the
clouds producing them are dense. The presence of weak absorption in
the $R$(3,3)$^l$ line at $\sim$ $-$205\,km\,s$^{-1}$ and $\sim$
$-$170\,km\,s$^{-1}$ toward stars $\alpha$ and $\alpha+$ (see Figure~7
of Paper~I) shows that the temperatures in these dense clouds are
higher than in clouds in the Galactic disk, but are
lower than 200\,K in the diffuse gas that permeates the
CMZ. Toward star $\beta$ the $R$(3,3)$^l$ high velocity absorption 
features are stronger relative to those in the  $R$(1,1)$^l$ profile, 
indicating somewhat higher temperatures in those parts of
the high velocity clouds. 

 Absorption by the $+$51\,km\,s$^{-1}$ cloud in the $R$(3,3)$^l$
line is completely absent (Paper~I, Fig.~7), demonstrating that
the temperature of that cloud is significantly below the mean
temperature of the diffuse gas in the CMZ. Although the 
 $+$51\,km\,s$^{-1}$ cloud probably lies within the CMZ we are 
 uncertain of its location. 

%------------------------------------------------------------
\subsubsection{The Absorptions near 0\,km\,s$^{-1}$ }
% subsubsec 4.7.2

We now turn to the unusually strong absorption features at
0\,km\,s$^{-1}$, which have full widths at zero intensity ranging from
30 to 50\,km\,s$^{-1}$ toward stars $\alpha$, $\alpha+$, and
$\beta$. This absorption feature toward Star $\alpha+$ in Figure  10 is 
the deepest single interstellar absorption feature in all 
H$_3^+$ spectra observed to date on any sightline. Toward Star $\alpha+$ 
and Star $\beta$, absorption by CO overtone band
lines is observed near 0\,km\,s$^{-1}$, implying that dense gas in the
local arm contributes to the absorption in the $R$(1,1)$^l$ line 
toward these two
stars. However, because strong absorption at 0\,km\,s$^{-1}$ is
present in the $R$(3,3)$^l$ line toward all three stars (Fig. 7 of Paper~I), it
is clear that large column densities of warm diffuse CMZ gas at very
low radial velocities, are also present on these two sightlines.

Surprisingly, no absorption near 0\,km\,s$^{-1}$ in the overtone band
lines of CO is detected toward Star $\alpha$ \citep[][see
also Paper~I]{geb10}. Thus there are no dense clouds in the local arm on the
sightline toward this star to contaminate the spectrum of the
$R$(1,1)$^l$ line at low radial velocities. Diffuse gas in the local spiral arm 
could be present on this sightline and the H$_3^+$ it might contain could 
produce an absorption feature at low velocity. However, the strong absorption
 in the $R$(3,3)$^l$ line at this velocity shows that the
deep absorption at 0\,km\,s$^{-1}$ in the $R$(1,1)$^l$ line must be produced largely,
 if not entirely by warm H$_3^+$ within the CMZ.
  
In summary, although in the central part  of the CMZ the absorption by 
H$_3^+$ in the warm and diffuse gas exists over a wide range of blueshifted 
radial velocities as high as $\sim$$-$150\,km\,s$^{-1}$,  at the western edge of the 
CMZ the warm diffuse gas exists only over a narrow range of radial velocities 
close to 0\,km\,s$^{-1}$.

%============================================================
\section{DYNAMICS AND MORPHOLOGY}

It was postulated in Paper~I, but not demonstrated in detail, that the
warm diffuse gas in the CMZ is undergoing a radial expansion that
originated near the center of the CMZ. In this section we examine that
proposition more thoroughly, drawing upon the velocity profiles of the
$R$(1,1)$^l$ line of H$_3^+$ presented in Section~4. We also compare
our results to the earlier studies of the motion, morphology, and
physical properties of the EMR by \citet{kai72} and \citet{sco72b}.

%------------------------------------------------------------
\subsection{The data} % subsection 5.1

The Galactic longitudes of the stars and ranges of radial velocities
in which the warm diffuse gas is present, based on the spectra of the
$R$(1,1)$^l$ line of H$_3^+$ presented in Section~4, are summarized in
Table~1. Most of the sightlines contain broad blueshifted ``troughs''
of absorption by this gas.  Most of these troughs have one
edge at large negative velocity, well in excess of
$-$100\,km\,s$^{-1}$, and the other edge at a velocity near
zero. There are some variations, however. The troughs of three 
stars (NHS 21, $\delta$ and $\theta$) are noncontinuous, but contain narrow gaps, which may indicate the existence of
voids of the warm and diffuse gas in parts of their sightlines or
simply non-continuous distributions of velocities. For some other sightlines, the edge of the trough near 0\,km\,s$^{-1}$ could not be clearly
located, because the trough merges with absorption by warm dense gas likely
located in the CND (GCIRS\,3 and GCIRS\,1W) and/or overlaps with
strong foreground absorption by cold gas in the local arm, as discussed in
Section~4. For those stars the upper limits to the troughs are shown
as $\sim$0\,km\,s$^{-1}$ in Table~1. For others the trough extends
slightly to positive radial velocities.
    
For four stars broad troughs are missing and the absorptions by warm
diffuse gas are narrow. For the three stars at the western edge of the
CMZ, $\alpha$, $\alpha+$, and $\beta$, this is understandable in our
interpretation that their sightlines are perpendicular to the motion of
the gas (Section~5.2.). Star $\gamma$ located midway between the
center of the CMZ and its western edge, shows only warm diffuse gas at
high negative velocities. In this case the explanation may be that
this star is located near the front surface of the CMZ, which, as
discussed in Section~5.2, is expanding radially at high velocity.

Finally, for Star $\iota$, the sightline which intersects the 
Sgr B molecular cloud complex, unlike any of the other
sightlines the  broad absorption seen toward it extends to high positive 
velocities. Because strong absorption at these velocities
is also present in spectra of the overtone CO lines (Figure~7), much of the warm 
gas in which these absorptions arise must be dense and be associated with Sgr B.
However, as noted in Section~4.4, warm diffuse gas is clearly present at high 
negative velocities and may extend to very low negative velocities. We conclude that 
this gas  is in the CMZ, but is in front of Sgr B and is not associated with it.

%============================================================
% Figure 11
%============================================================

\begin{figure*}
\includegraphics[angle=-0,width=0.98\textwidth]{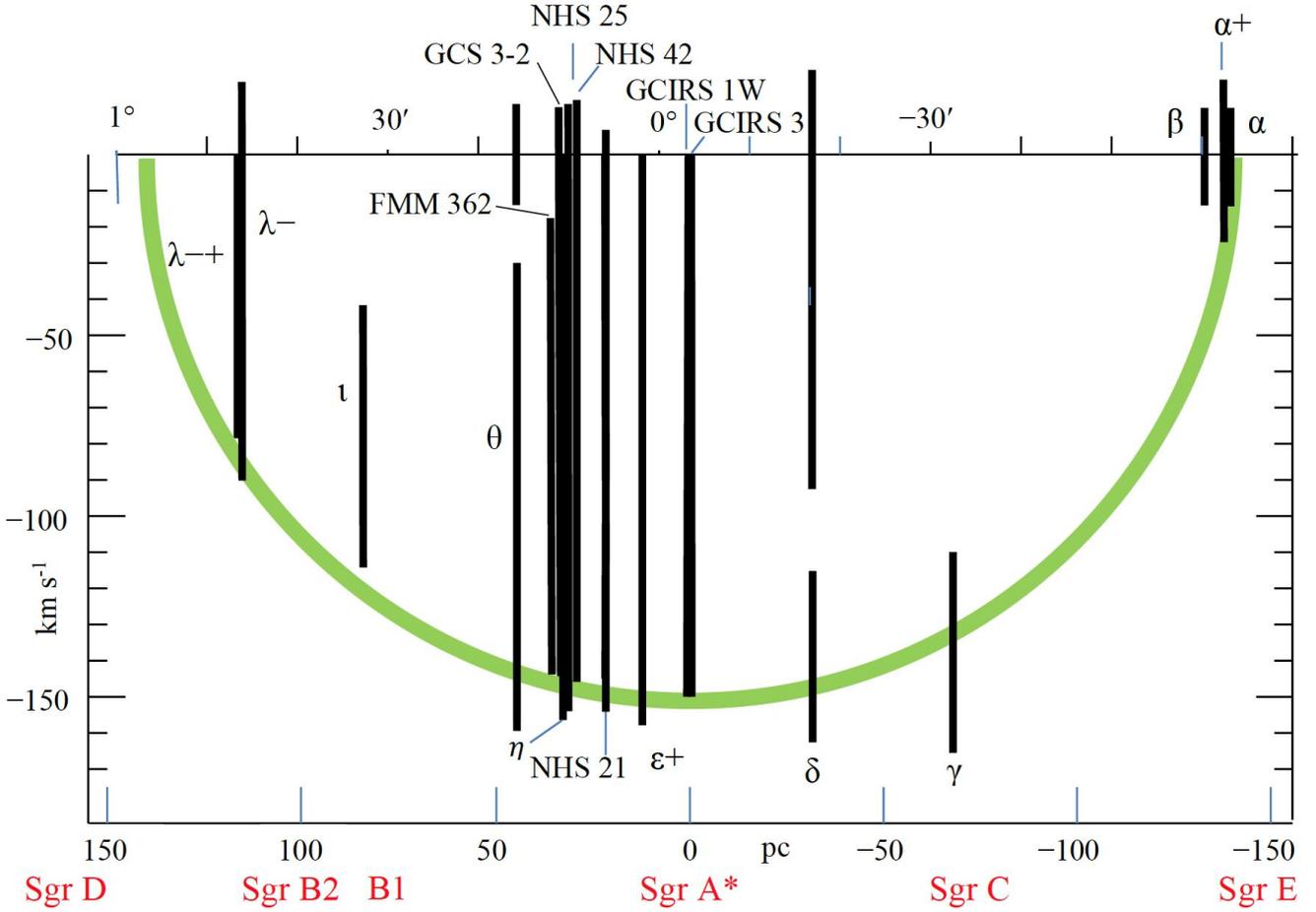}
\caption{The ($l$, $v$) diagram for the observed velocity profiles
of the H$_3^+$ toward 18 stars, created using the data compiled 
in Table~1. Galactic longitude is given at top and
longitudinal distance from Sgr~A$^\ast$ (at $G_{\rm lon} = -0.056\degr$) 
in pc is at bottom. Approximate $G_{\rm lon}$ and distances of
Sagittarius radio sources with respect to Sgr~A$^\ast$: Sgr~E
($-$1.079\degr, $-$143\,pc); Sgr~C ($–$0.571\degr; $-$72\,pc); Sgr~B1
(0.506\degr, 79\,pc), Sgr~B2 (0.667\degr, 101\,pc), and Sgr~D
(1.126\degr, 166\,pc) are shown at bottom. The semicircle indicates
the front of the expanding gas with radius 1\degr~and velocity of
150\,km\,s$^{-1}$. The black vertical line segments indicate the
ranges of radial velocities for which absorption by warm diffuse gas
is detected. Individual sightlines are labeled. For Star $\iota$ note
the discussion at the end of Section~4.4.}
\end{figure*}

%============================================================
% Figure 12
%============================================================
\begin{figure*}
\begin{center}
\includegraphics[scale=.61]{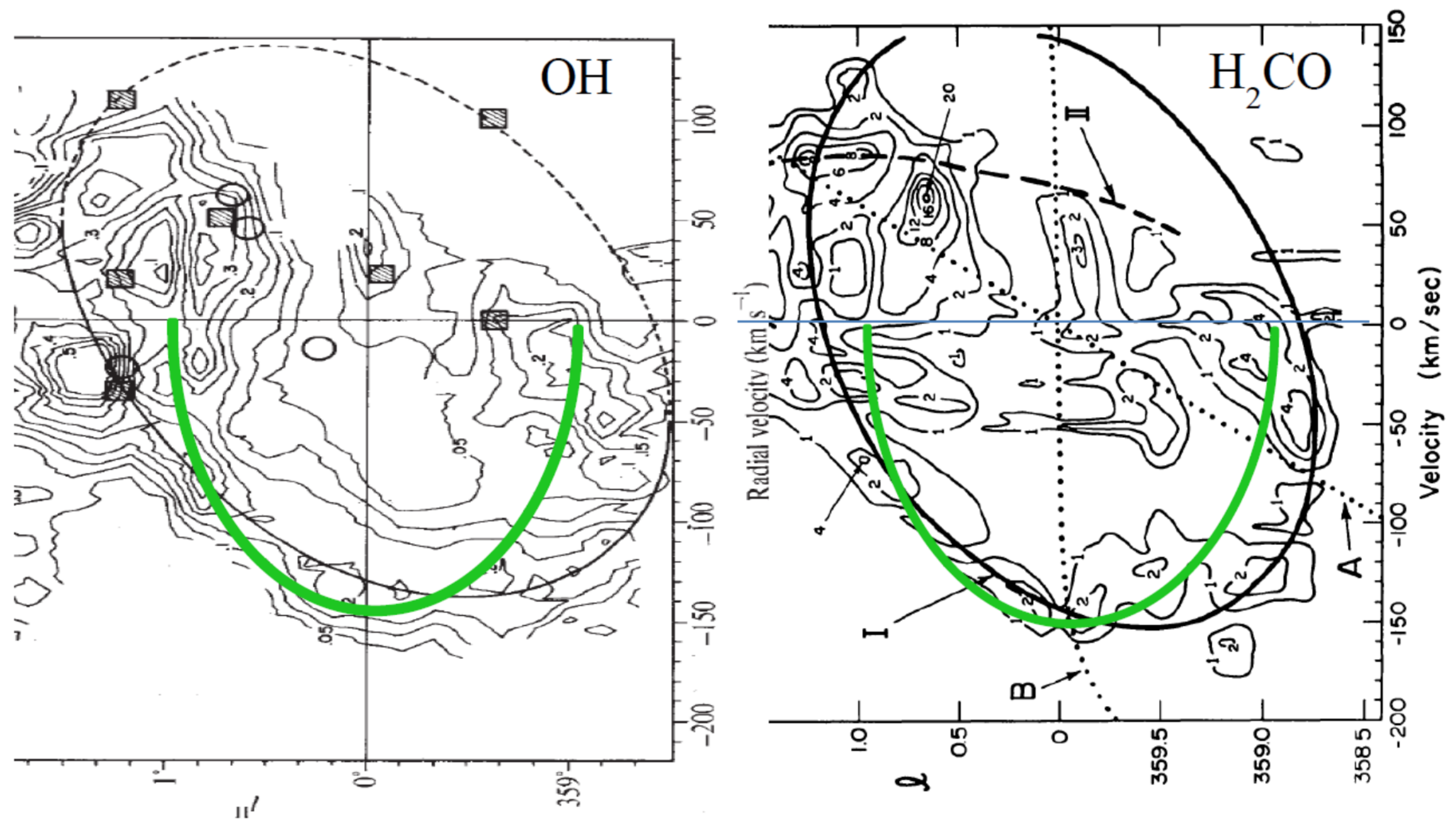}
\end{center}
\caption{Comparisons between the ($l$, $v$) diagram of the front
surface of the expanding diffuse gas obtained from H$_3^+$ (green
lines) and the those of the EMR by \citet{kai72} (left panel,
continuous line) and \citet{sco72b} (right panel, continuous
line). Kaifu et al. used the contour map of the 1667\,MHz OH
absorption reported by \citet{mcg70} to draw the front half of the EMR
and two 1.3\,cm NH$_3$ emission lines (shaded rectangles) to fit the
back part of the EMR (left panel, dashed line). Scoville's ($l$, $v$)
diagram is based on an absorption line of H$_2$CO.}
\end{figure*}
%============================================================

%------------------------------------------------------------
\subsection{The ($l$, $v$) diagram of the warm diffuse CMZ gas}
% subsection 5.2

The data in Table~1 are plotted in the ($l$, $v$) diagram shown in
Figure~11. The green semicircle of radius 1\degr~and 150\,km\,s$^{-1}$
indicates the outer edge of the warm ($T\sim 200$\,K) and diffuse ($n
\sim 50$\,cm$^{-3}$) CMZ gas, assumed to be expanding radially at
150\,km\,s$^{-1}$, approximately at the same velocity as the EMR.

As can be seen in the figure, the highest observed negative radial
velocities of the warm diffuse gas match this simple model very
well. The largest deviations are somewhat to the west of center, where the 
sightlines to Stars $\gamma$ and $\delta$ contain gas at moderately higher 
maximum negative velocities than the model. These might be explained 
by asymmetries in the initial ejection of the gas or in preexisting density
nonuifornities in the CMZ.

The close match between the highest negative velocities in the  H$_3^+$
absorption troughs and the velocities observed in the EMR
strongly indicates that the warm diffuse gas observed at
the highest velocities is located at or close to the EMR, lending
strong support to an interpretation of the spectra as the result of
radial expansion. The warm diffuse gas observed at lower
velocities on these sightlines would then lie inside the CMZ (as is 
suggested by the velocity profile of Star $\gamma$ in Figure 9). This
interior gas must also be undergoing largely outward radial motion. Little 
or none of it is falling toward the center (a possibility that might have been expected 
because of the gravitational potential; see Section 6.3). If some of it were infalling,  
absorption would also be expected to be present over an extensive range 
of positive velocities. Such absorption is not observed. Finally and most 
critically, the considerably lower radial velocities observed toward the 
two stars near the eastern edge of the CMZ, and in
particular the very low radial velocities and narrow velocity ranges
seen on sightlines to the three stars close to the western edge of the
CMZ, are key support for the radial expansion interpretation. This
is because (1) radial expansion viewed at the edges of the CMZ is
perpendicular to the line of sight, and (2) the strengths of these
absorptions imply column lengths of tens of pc (Paper~I).

Note that apart from the velocity profile toward Star $\iota$, none
of the spectra of H$_3^+$ lines reveals warm diffuse highly redshifted gas. In a
symmetric radial expansion such gas should be present if the
background stars are spread throughout the front and rear halves of
the CMZ. As discussed in Paper~I, we believe that the absence of
redshifted H$_3^+$ is a selection effect, largely the result of the
brightness constraint imposed in order to obtained spectra of H$_3^+$
of sufficient quality. The existence of highly redshifted gas in the
EMR is demonstrated by the radio observations of \citet{kai74} and
\citet{sco72b}.
 
%------------------------------------------------------------

\subsection{Comparison with ($l$, $v$) diagrams of Kaifu et al. and Scoville}
% subsection 5.3

In Figure~12 the ($l$, $v$) diagrams of the EMR obtained from radio
absorption spectra of OH by \citet{kai72} and of H$_2$CO by
\citet{sco72b} are compared with the simple ($l$, $v$) diagram based
on the highest observed velocities of H$_3^+$ (Figure~11). In view of
the simplicity of the fit to the H$_3^+$ data, based on widely spaced data
points, and the low resolution of the radio data, the infrared and radio
diagrams are in good agreement. The ($l$, $v$) diagrams and properties
of the EMR are discussed in more detail below.

%------------------------------------------------------------
\subsubsection{Dense or diffuse gas}
% subsubsec 5.3.1 

While the expanding gas observed in H$_3^+$ has a low mean density
($n\sim 50$\,cm$^{-3}$, Paper~I), both \citet{kai72} and \citet{sco72b} 
reported the gas forming the EMR to be dense ($n \geq
10^3$\, cm$^{-3}$). Kaifu et al. used the ($l$, $v$) diagram of the
strongest hyperfine component of the $\Lambda = 3/2$~OH
$\Lambda$-doubling transition at 1667.36\,MHz reported by \citet{mcg70}
to draw the front half of the EMR and used observations of
radio NH$_3$ emission lines at two locations to aid in filling in the
rear half. From the NH$_3$ emission they concluded the gas density in
the EMR to be 10$^3$--10$^4$\,cm$^{-3}$. The two data points of 
NH$_3$ (see Figure~12 left) may well be coincident with the EMR, but
we believe that determining the density of the EMR from NH$_3$
emission is not justified. The OH lines are more likely to result from
the absorption of scattered microwave radiation from the Galactic
center than from the cosmic microwave background. Because of the large
beam size \citep[12\farcm2 at the 210\,ft Parkes telescope,][]{mcg70}
detailed analysis of the OH data is a complex and nearly impossible
task. Nevertheless, that the 18\,cm line is in absorption suggests
much lower densities than assumed by \citet{kai72}.

% nH2 > 10^3->  nH2 > 10^3 cm-3

The ($l$, $v$) diagram proposed by Scoville (1972) is based on the
radio absorption spectrum of the H$_2$CO 1$_{10}$ $\leftarrow$
1$_{11}$ $K$-doubling transition at 4829.73\,MHz (6.2\,cm). However,
the density of the EMR, $n_{\rm H_2} > 10^3$\,cm$^{-3}$ adopted by him
is based on the $J=1\rightarrow 0$ CO line emission. CO emission in
this line is ubiquitous in the CMZ and we believe that an estimation
of the density of gas in the EMR from it is not justified. The
excitation of the energy levels of the $K$-doublet of H$_2$CO is a
complex problem \citep{tow69,oka70} and the
determination of cloud density based on its absorption spectrum is
even more complicated than that for OH.

From these considerations and the good agreement of the ($l$, $v$)
diagram for the front of the expanding gas and the EMR, we believe
that the gas in the EMR, which is located at the outer edge of the
CMZ, has similar properties to the diffuse gas observed in H$_3^+$
throughout the CMZ and that they are physically connected. We note
that in the spectra published in Paper~I there are a few sightlines
where weak absorption by overtone lines of CO is present at negative velocities near
$-$150\,km\,s$^{-1}$, suggesting that there is some dense molecular
gas in the CMZ that is near or part of the EMR. That would not be
surprising, as high velocity gas within the CMZ should be impacting
the EMR and driving it outward. In addition the EMR must be colliding
with and accumulating gas from the external interstellar medium. Both
of these processes will tend to increase the density where the impacts
are occurring and thus it is conceivable that the EMR contains regions
of dense molecular gas. However, based on spectroscopy of H$_3^+$ and
the CO overtone band lines, these regions appear to be few and far
between. Reducing the density of the gas in the EMR by 1--2 order of
magnitude, as suggested above, brings the kinetic energy of the EMR
reported by \citet{kai72} (10$^{55}$--10$^{56}$\,erg) and
Scoville (1972) ($> 6 \times 10^{53}$\,erg) into better agreement
with the kinetic energy of the EMS
($2 \times 10^{54}$\,erg) estimated by \citet{sof95b}.

% Expanding Molecular Shell =>  Expanding Molecular Ring?

%------------------------------------------------------------
\subsubsection{Ring, filled disk, or both?}
%subsubsec 5.3.2

Both the high column densities of H$_3^+$ and the broad absorption
troughs of its $R$(1,1)$^l$ and $R$(3,3)$^l$ lines on many sightlines,
as illustrated in Figure~11, lead to the conclusion in Paper~I that
warm diffuse gas fills much of the CMZ. In contrast, the ($l$, $v$)
diagrams of OH and H$_2$CO indicate rings of molecular gas with a void
in the center. Although quantitative analysis is complicated for OH
and H$_2$CO and is beyond the scope of this paper, we suspect that 
the differences in the distributions of these species are related to differences in
their chemistries. The molecular ion H$_3^+$, being extremely active
chemically, has a very short lifetime in the CMZ, where electrons
abound due to rapid ionization of H and H$_2$ by the high cosmic ray
flux. From the high rate constant for dissociative recombination
$\sim$ $10^{-7}$\,cm$^3$\,s$^{-1}$ at $T\sim 200$\,K (McCall et
al. 2004) and the high electron number density $n_{\rm e}\sim$
$0.33$\,cm$^{-3}$ (Paper~I), the lifetime of an H$_3^+$ ion in the CMZ
is estimated to be on the order of years. Thus, H$_3^+$ is constantly
being created and destroyed throughout the CMZ and should be
distributed more or less uniformly in it.

On the other hand, OH and H$_2$CO have much longer lives and their
distributions are more specific to the environment. They have 1000
times higher rates of photodissociation by FUV than H$_3^+$\citep{dis88,hea17}. 
The UV field has been estimated to be 1,000 times higher in the CMZ 
than in the solar neighborhood \citep{rod04}. In the CMZ most of the stars,
including many of the hottest and most luminous ones, e.g., those in
the Central, Quintuplet, and Arches clusters, are centrally
concentrated. Hence, the UV intensity should decrease with
distance from the center beyond $\sim$30\,pc, apart from local
effects. This may result in the abundances of OH and H$_2$CO increasing
with increasing distance from the center, showing peaks at the EMR, and
thus appearing as rings. Readers are also referred to the ($l$, $v$)
diagram of OH absorption in Fig.~9 of \citet{coh76}, 
where the filling inside the EMR for this molecule seems somewhat 
more uniform than in Figure~12 (left panel).

%------------------------------------------------------------
\subsubsection{Rotation?}
%subsubsec  5.3.3.
  
While the ($l$, $v$) diagram of H$_3^+$ is interpreted as due to
purely expanding gas in this paper, both \citet{kai72} and \citet{sco72b} 
interpreted their ($l$, $v$) diagrams as indicating
an {\it expanding and rotating ring} with expansion and rotational velocities 
130$\pm$5\,km\,s$^{-1}$ and 50$\pm$20\,km\,s$^{-1}$, and
145\,km\,s$^{-1}$ and 50\,km\,s$^{-1}$, respectively. The difference
between their estimates and ours stems from the difference in
symmetries with respect to $G_{\rm lon}$. In Figure 12, the (half) ellipse drawn as
showing the front of the expanding gas based on spectra of H$_3^+$ is
symmetric with respect to longitudinal distance from Sgr~A$^\ast$ while
the elliptical fits to the ($l$, $v$) diagrams of OH and H$_2$CO are
not. However, in our examination of the contour plots
of these two molecules without the elliptical fits, we do not see
convincing evidence for this asymmetry.

The essential difference between the half ellipse obtained from
H$_3^+$ and the ellipses derived for the EMR is that while the former
gives zero radial velocity at the westmost edge, the latter extend
farther to negative velocities to the west because of the asymmetry
with respect to $G_{\rm lon}$. If the latter were correct, the OH
ellipse implies that an absorption trough of H$_3^+$ would be present
from 0 to $\sim$ $-$80\,km\,s$^{-1}$ in the spectra toward Stars
$\alpha$, $\alpha+$, and $\beta$; the H$_2$CO ellipse would predict an
absorption trough from 0 to $\sim$ $-$120\,km\,s$^{-1}$. Such troughs
are not seen in H$_3^+$, demonstrating that the asymmetry does not
exist and that the EMR is not rotating.

%============================================================
\section{ORIGIN AND FUTURE OF THE RADIAL EXPANDING GAS}
% section 6.  

The concerted gas motion away from the center discovered by \citet{kai72} and \citet{sco72b}
and confirmed in this paper must be
either the result of a large accumulation of energetic stellar events
or the result of one or more much more energetic expulsion events near
the center in the recent past, possibly associated with
Sgr~A$^\ast$. In the following sections we attempt to constrain the
possible explanations of the observed radial expansion and to predict
the future of the CMZ's diffuse gas.

%------------------------------------------------------------
\subsection{Momentum considerations}
% subsec 6.1

Using the estimated mean density and filling factor of the warm diffuse
gas, the total mass of the warm diffuse expanding gas was calculated in
Paper I to be $6\times 10^6 M_\odot$ (see Paper I, Table~7).  For
comparison, the mass of the EMR had been estimated to be $3\times 10^6
M_\odot$ by \citet{sco72b}  and $1\times 10^7 M_\odot$ by \citet{kai74}. 
However, as discussed in Section~5.2.1 we believe their
estimates of density are too high by 1--2 orders of magnitude.

We estimate that the current radial momentum of the CMZ's warm diffuse
gas to be approximately $5 \times 10^8 M_\odot$\,km\,s$^{-1}$ (using a
mean gas velocity $v$ of 75\,km\,s$^{-1}$). This is comparable to that
generated by approximately 10$^4$ core-collapse supernovae, assuming
that each supernova ejects $10 M_\odot$ at an average speed of
3,000\,km\,s$^{-1}$. Note that this estimate is a lower limit, because
the radial momentum of the CMZ gas decreases with time as a
consequence of the gravitational attraction by the enclosed mass, as
discussed below. The current kinetic energy of this gas (assuming a
mean $v^2$ of $10^4$\,km$^2$\,s$^{-2}$) is $\sim 5 \times
10^{53}$\,erg.

This decrease in radial momentum is significant. A crude
characteristic expansion time of the CMZ gas, assuming the current
maximum velocity of 150\,km\,s$^{-1}$ and the CMZ radius of 150\,pc,
is $\sim 1\times 10^6$\,yr. Using Fig.~14 of \citet{sof13}, \citet{geb19b}
estimate that the deceleration due to gravity from a
distance of 30\,pc from Sgr~A$^\ast$ to the edge of the CMZ at
$r=150$\,pc is $\sim 1 \times 10^{-6}$\,cm\,s$^{-2}$. Because of the
gravitational potential, in one million years gas ejected into the CMZ
near its center would have decelerated by a few hundred km\,s$^{-1}$
from an initially much higher velocity than the maximum velocity of
150\,km\,s$^{-1}$ presently observed. This implies that the current
radial momentum in the outflowing gas is considerably less than half
of what it was one million years ago.  Using a shorter characteristic
time to account for the deceleration, would still require a much
higher initial radial momentum than the above value.
  
The supernova rate in the CMZ has been variously estimated to be
(1–-10) $\times 10^{-4}$\,yr$^{-1}$ \citep{cro11, zub13}.
Even the highest estimated rate would fall short of
generating the current radial momentum by an order of magnitude in one
million years, and by more if the larger initial radial momentum
required by a shorter characteristic time are used.

The radial momentum of the CMZ gas also exceeds, by a much larger
factor, the total radial momentum generated by winds of all of the
several hundred massive stars of the three massive clusters (Central,
the Quintuplet, and the Arches) in the central 30\,pc of the CMZ
during their few million year lifetimes. Assuming each star ejects $10
M_\odot$ of its mass at 1,000\,km\,s$^{-1}$ before exploding, $3\times
10^4$ such stars would be required to account for the observed radial
momentum, two orders of magnitude more than are currently
present. Although estimates are highly uncertain, normal red giants
might have contributed an appreciable fraction of the current mass and
radial momentum of the CMZ, but they cannot be responsible for the
high velocities that are observed for much of this gas.

The births of the above three massive clusters, must have been
accompanied with ejection of large mass with high velocities. Their
time of formation 4.8 million years ago for the Quintuplet Cluster
\citep[e.g.][]{sch14} and 2.5 million years ago for the Arches
Cluster \citep[e.g][]{esp09}, however, are much greater than
the characteristic expansion time of the CMZ and EMR gases.

%------------------------------------------------------------
\subsection{Past explosive events}

In view of the above discussion, one must consider whether the
expanding gas was produced by some other means. Possibly, the gas is
the aftermath of one or more explosive events in the nucleus
associated with the supermassive black hole. In view of the above
estimates of the characteristic time such events would have occurred
several hundred thousand to one million years ago. Both \citet{kai72} 
and \citet{sco72b} invoked such events to account for the
EMR.

% x-ray / X-ray => X-ray

Expulsion events have been previously suggested by many authors. In a
series of papers on north and south radio spurs beginning with \citet{sof73},
Sofue proposed that the spurs are results of star bursts near
the GC, contrary to the consensus at the time that they are supernova
remnants in the vicinity of the solar system. His magnetohydrodynamic
wave calculation \citep{sof76, sof77} based on \citet{uch70} provided
support for the idea. Observations of radio continuum emission by Sofue
and Handa (1984) revealed a giant Q-shaped radio loop, the Galactic
center lobe (GCL). This result together with X-ray maps by \citet{mcc83} and 
\citet{mcc90} led \citet{sof94} to propose
that a giant explosion at the GC, of energy $3 \times 10^{56}$\,erg,
occurred 15 million years ago. From further studies of X-rays via
ROSAT all-sky maps \citep{sno97} and adiabatic shock wave
calculations based on \citet{sak71} and \citet{mol76}, \citet{sof00}
proposed a bipolar hypershell model for the ejecta. This model
can be considered as anticipating the discovery of the bipolar
$\gamma$-ray Fermi Bubbles \citep{dob10, su10}. Although 
estimated ages of the Fermi Bubbles vary by
large factors, they generally suggest an explosion at the GC about 10
million years ago. The relation between microwave lobes, X-ray and
$\gamma$-ray maps, all indicating explosive events in the GC, are
discussed in \citet{kat18} and \citet{sof19}. These events
appear to be more than than 10 times older and possessing much higher
energy than could account for the current state of the warm diffuse
CMZ gas reported here.

% Finkenbeiner? Mo?

The 430\,pc bipolar radio bubble recently detected by MeerKAT
 \citep{hey19} indicates explosive events of
much smaller scale both in time (a few million years) and total energy
($7 \times 10^{52}$\,erg). However, the gas velocity of
30\,km\,s$^{-1}$ \citep{law09} used for estimating the time may
well be much larger. If so the time since the event occurred is under one million 
years in agreement with estimate in this paper. Indeed the value of ``a few times 
10$^5$--10$^6$ years ago'' given by \citet{yus19a} contains an estimate of 
somewhat less than 1 million years, well within our proposed range of ages. 
The energy of the expanding CMZ gas of $5 \times 10^{53}$\,erg is an order 
of magnitude higher than the estimated energy of the radio lobe. Together 
with the question of why the explosion energy is in-plane, further discussion of 
these issues is left for the future. From the agreement of the ages of these events, 
it is possible that the expanding gas observed in H$_3^+$ and the MeerKAT
radio bubble were caused by the same explosive event.

%------------------------------------------------------------
\subsection{Deceleration and future infall of the CMZ gas}

Currently, there is no evidence that at the present time the expanding
diffuse gas that the CMZ contains is being actively driven outward. It
is thus of interest to consider the effect of gravity on the gas. The
deceleration by the gravitational force from the enclosed mass of the
Galaxy \citep[][ Fig.~14]{sof13} increases rapidly from 30\,pc from
the center inward, but is fairly constant at approximately $1 \times
10^{-6}$\,cm\,s$^{-2}$ between 30\,pc and 150\,pc from Sgr~A$^\ast$. As
discussed earlier, the decrease in radial velocity over the
characteristic time of $1 \times 10^6$\,yr for the expanding gas, is
then approximately 300\,km\,s$^{-1}$. This is roughly twice the
observed current maximum expansion velocity of the warm diffuse
gas. If the characteristic time is roughly half of the above value,
due to taking into account the previous deceleration of the gas, the
decrease in velocity is equal to the current maximum expansion
velocity (See Appendix for a formal calculation).

As can be seen in Fig.~14 of \cite{sof13}), beyond the radius
of the CMZ, the enclosed mass increases slightly less rapidly with
distance from the center than at smaller radii. However at 300\,pc the
deceleration due to gravity has only decreased to two-thirds the above
value. Thus it is clear that the expanding gas in the EMR and CMZ will
not escape from the Galactic center. A more quantitative calculation
given in the Appendix leads us to the same conclusion.

The above considerations suggest the following scenario. The expansion 
of highest velocity warm diffuse gas will end in roughly $5 \times
10^5$\,yr. Outward motion of the lower velocity gas in the interior of
the CMZ will end sooner. Warm diffuse gas will begin to fall back
towards the center. Gas densities in the inner part of the CMZ will
increase and new episodes of star formation within the CMZ as well as
outbursts associated with gas accreting onto Sgr~A$^\ast$ could
follow. These events could begin to occur after an additional few
hundred thousand years.

Note that this scenario ignores any role played by dense molecular
clouds, whose total mass has been estimated to be several times that
of the diffuse CMZ gas \citep{sof17}.  Because the filling factor in the 
CMZ of the dense gas is believed to be much smaller than that of the 
diffuse gas, the evolution of the diffuse gas may be largely independent 
of the dense clouds, unless one of the latter falls into the center at an
earlier time.

%============================================================
\section{SUMMARY AND CONCLUSIONS}
% section 7

We have obtained velocity profiles of the $R$(1,1)$^l$ line of H$_3^+$
at 3.7\,$\mu$m toward 29 stars whose sightlines are toward the CMZ of
the Galaxy and extend in longitude nearly fully across the CMZ; we
present 18 of them in this paper. These18 stars are sufficiently deeply
embedded in the CMZ that strong absorptions are observed in the
$R$(1,1)$^l$ line, implying long columns of H$_3^+$. As discussed in
Paper~I, they, together with supporting observations of other lines of
H$_3^+$, imply that warm ($\sim$200 K) and diffuse
($\sim$50\,cm$^{-3}$) gas fills the majority of the volume of the CMZ.

The absorption lines of H$_3^+$ seen toward those stars in the central
part of the CMZ are broad and blueshifted, extending from $\sim$ $-$150\,km\,s$^{-1}$
to $\sim$ 0\,km\,s$^{-1}$. Toward most of the stars the absorptions are
continuous and form troughs, while toward a few they are
discontinuous indicating patchy voids in the diffuse gas. In general
the maximum (negative) velocities of the line profiles decrease with
distance of the sightline from the center of the CMZ. Sightlines at
the western edge of the CMZ show narrow absorptions centered at
0\,km\,s$^{-1}$. No sightlines are available at the eastern edge but
those closest to it have considerably narrower absorption troughs and
considerably lower maximum (negative) absorption velocities than the
lines on more centrally located sightlines.

The ($l$, $v$) diagram of these line profiles are well fitted by a
semicircle whose radius corresponds to the maximum velocity of
150\,km\,s$^{-1}$ and whose linear dimension is that of the CMZ. The diagram
implies that the warm diffuse gas is expanding from an origin near the
center of the CMZ. This motion is similar to that of the Expanding
Molecular Ring (EMR) discovered by \citet{kai72} and \citet{sco72b}, 
although there are three qualitative differences:

\begin{enumerate}
\item The EMR was originally interpreted as composed of dense ($n>
  10^3$\,cm$^{-3}$) gas while the gas probed by H$_3^+$ is diffuse
  ($n\sim$50\,cm$^{-3}$). We believe that the density estimates of \citet{kai72} and \citet{sco72b}, 
  which were not directly obtained from the observed absorption spectra of OH and
  H$_2$CO, but from emission spectra of NH$_3$ and CO, respectively,
  are overestimates and that the EMR has roughly the same density as is
  derived from observations of H$_3^+$.

\item While the EMR as viewed in OH and H$_2$CO is a ring with a void at
the center, the gas probed by H$_3^+$ fills the CMZ. While H$_3^+$ is
easily understood as being present throughout the CMZ, the intense UV
radiation field in the central part of the CMZ could account for the abundances of 
OH and H$_{2}$CO increasing with distance from the center and reaching maxima at the EMR.

\item The EMR was reported to have a rotational component with velocity
$\sim$50\,km\,s$^{-1}$ both by \citet{kai72} and \citet{sco72b}. We believe
that their arguments in support of this motion are not convincing. The
observations of H$_3^+$ toward the edges of the CMZ are unequivocal in
demonstrating very little or no rotational component.
\end{enumerate}
  
\noindent
Analysis of the spectra of H$_3^+$ allow us to infer that a face-on
view of the CMZ would show it to be circular. This revives the circular
face-on view of the CMZ proposed by \citet{kai72} and \citet{sco72b},  
which fell out of favor after the paper by \citet{bin91}. Although 
there is good evidence for elliptical orbits with
high eccentricity on scales of a few kpc, the application of the
Binney et al. theory to the CMZ, which is more than 10 times smaller
and for which there is no evidence for the presence of a barred potential is not
justified. We believe that the face-on views of the CMZ as an ellipse
with high eccentricity, such as those in Fig.~2 of \citet{rod06}, Fig.~21 
of \citet{bal10}, and Fig. 4 of \citet{tsu18} are incorrect.

The observations of H$_3^+$, together with those of OH and H$_2$CO in the
EMR indicate that nearly $10^7 M_\odot$ of gas was expelled from
central region of the CMZ roughly 600,000 years ago (see
Appendix). The current kinetic energy and momentum in the gas greatly
exceed what could be produced by supernovae and/or stellar winds, and
the ejection is probably related to activity of Sgr~A$^\ast$. Compared
with explosions inferred from other observations like the microwave lobes
\citep{sof84}, X-ray maps \citep{mcc90, sno97}, and the $\gamma$-ray 
Fermi bubbles \citep{su10}, both the
time and energy are an order of magnitude less. The time scale agrees
with that of the recently discovered 430\,pc microwave bubble 
\citep{hey19, yus19a, yus19b}. The origin of the
expansion of the diffuse gas probed by H$_3^+$ may be the result of
the same event or events that created the microwave bubble.

Due to the significant deceleration of the gas produced by the
enclosed Galactic mass, it can be deduced that originally the 
maximum expansion velocity was considerably higher than 
the current value of $\sim$150\,km~s$^{-1}$ .
The ongoing deceleration will halt the expansion of the EMR
within $\sim 5 \times 10^5$\,yr and may have already nearly halted the
expansion of some of the diffuse gas interior to the EMR. This 
will result in infall of the gas (unless it is prevented by
another outburst in the center), causing new episodes of star formation
and violent events associated with accretion onto Sgr~A$^\ast$.

%============================================================
\acknowledgments

We are grateful to F. Yusef-Zadeh, Y. Sofue, H. Liszt, and T. Oka for
many helpful discussions over the last 20 years, as well as comments on an earlier 
draft of this paper. We thank N. Indriolo and M. Goto for helping to obtain and reduce 
some of the spectra, and M. Goto for other assistance. We also are indebted to the referee 
for several helpful suggestions This paper is based in large part on 
observations obtained at the international Gemini Observatory (Programs GS-2003A-Q-33, 
GS-2008A-C-2, GS-2009A-C-6, GN-2010AQ- 92, GN-2011A-Q-105, GN-2011B-Q-12, 
GN-2011B-Q-90, GN-2012A-Q-75, GN-2012A-Q-121, GN-2013A-Q-114, GN-2014A-Q-108, 
GS-2014A-Q-95, GN-2015A-Q-402, GS-2015A-Q-96, GN-2016A-Q-96, GS-2016A-Q-102, and
GS-2017A-Q-95). Gemini Observatory is a program of NOIRLab,  which is managed by the 
Association of Universities for Research in Astronomy (AURA) under a cooperative agreement 
with the National Science Foundation. on behalf of the Gemini Observatory partnership: the National 
Science Foundation (United States), National Research Council (Canada), Agencia Nacional de 
Investigaci\'{o}n y Desarrollo (Chile), Ministerio de Ciencia, Tecnolog\'{i}a e Innovaci\'{o}n (Argentina), 
Minist\'{e}rio da Ci\^{e}ncia, Tecnologia, Inova\c{c}\~{o}es e Comunica\c{c}\~{o}es (Brazil), and 
Korea Astronomy and Space Science Institute (Republic of Korea).
We thank the staffs of the Gemini Observatory, the Very Large Telescope, the Subaru 
Telescope, and the United Kingdom Infrared Telescope for their support.

\facilities{Gemini:Gillett, Gemini:South, Subaru, UKIRT}

%============================================================
\clearpage
\appendix
\section{Gas under constant deceleration}

The gravitational deceleration of the spherically expanding gas in the
spherically distributed mass of the CMZ is

\begin{equation}
\frac{d^2r}{dt^2} = -\frac{GM(r)}{r^2},
\end{equation}
% (A.1)
\noindent
% mas -> Mass
where $r$ is the distance from the center of expansion, $G$ is the
gravitational constant, and $M(r)$ is the mass within the sphere of
radius $r$. We use $M(r)$ given in Fig.~14 of \citet{sof13} which is
approximated as $M(r) = \mu r^2$ between 30\,pc $\equiv r_0$ and
150\,pc $\equiv R$ with $M(r_0) = 6.0 \times 10^7 M_\odot$ and $M(R) =
1.5 \times 10^9 M_\odot$.  $G\mu \equiv \gamma = 1.0 \times
10^{-6}$\,cm\,s$^{-2}$. For this mass distribution, the deceleration
is constant between 30\,pc and 150\,pc,

\begin{equation}
  \frac{d^2r}{dt^2} = –\gamma,\\
       {\rm ~i.e.,~~}\\
       \frac{dr}{dt} = v = v_0 – \gamma t,\\
            {\rm ~and~~} \\
        r = r_0 + v_0 t – \gamma \frac{t^2}{2},          
\end{equation}
%   (A.2)
\noindent
where $v_0$ is the velocity of the gas at 30\,pc and the time $t$ is
set to be 0 at 30\,pc. These quadratic simultaneous equations with $t$
and $v_0$ as unknowns give the solution

\begin{equation}
  t = \left[\left(\frac{v}{\gamma}\right)^2 +
  \frac{2 (r – r_0)}{\gamma} \right]^{1/2} -
    \frac{v}{\gamma},\\
    {\rm ~and~~} \\
    v_0 = v + \gamma t.
\end{equation}
%    (A.3)
\noindent
For the gas at the front of expansion with $v = V = 150$\,km\,s$^{-1}$
and $r – r_0 = R – r_0 = 120$\,pc the time of expansion $T$ is
calculated to be
\begin{equation}
  T = 5.3 \times 10^5\, {\rm year,~}\\
  {\rm and~the~initial~velocity~} \\
  v_0 = 310\,{\rm km\,s^{-1}},
\end{equation}
% suggestion T -> t_{\rm exp}. T sounds like temperature.
\noindent
Thus the gas at the front of the expansion had a speed of
310\,km\,s$^{-1}$ at 30\,pc and has decelerated to 150\,km\,s$^{-1}$
during the expansion from 30\,pc to 150\,pc. The expansion from the
origin to 30\,pc must have happened in much shorter time than $T$. We
thus conclude that the event which initiated the expansion of the gas
occurred about $6 \times 10^5$\,years ago.

Figure~14 of \citet{sof13} indicates that the relation $M(r) = \mu r^2$
holds approximately beyond $r = 150$\,pc up to 300\,pc. Because of the
constant deceleration in the region, the gas will reverse its velocity
and fall back to the center long before it reaches 300\,pc and join the
inevitable inflow toward the center \citep{mor96}.

%===========================================================
% \vspace{10mm}

%============================================================
\end{document}